# Machine Learning in Thermodynamics: Prediction of Activity Coefficients by Matrix Completion

*Fabian Jirasek[1,2†*], Rodrigo A. S. Alves[3†], Julie Damay[4†], Robert A. Vandermeulen[3], Robert Bamler[1], Michael Bortz[4‡], Stephan Mandt[1‡], Marius Kloft[3‡], Hans Hasse[2‡]*

[1]Department of Computer Science, University of California, Irvine, USA

[2]Laboratory of Engineering Thermodynamics (LTD), TU Kaiserslautern , Germany

[3]Machine Learning Group, Department of Computer Science, TU Kaiserslautern, Germany

[4]Fraunhofer Institute for Industrial Mathematics ITWM, Kaiserslautern, Germany

[†]These authors contributed equally to this work.

[‡]These authors jointly directed this work.

*Correspondence to: fabian.jirasek@mv.uni-kl.de

**Abstract**

Activity coefficients, which are a measure of the non-ideality of liquid mixtures, are a key property in chemical engineering with relevance to modeling chemical and phase equilibria as well as transport processes. Although experimental data on thousands of binary mixtures are available, prediction methods are needed to calculate the activity coefficients in many relevant mixtures that have not been explored to-date. In this report, we propose a probabilistic matrix factorization model for predicting the activity coefficients in arbitrary binary mixtures. Although no physical descriptors for the considered components were used, our method outperforms the state-of-the-art method that has been refined over three decades while requiring much less training effort. This opens perspectives to novel methods for predicting physico-chemical properties of binary mixtures with the potential to revolutionize modeling and simulation in chemical engineering.

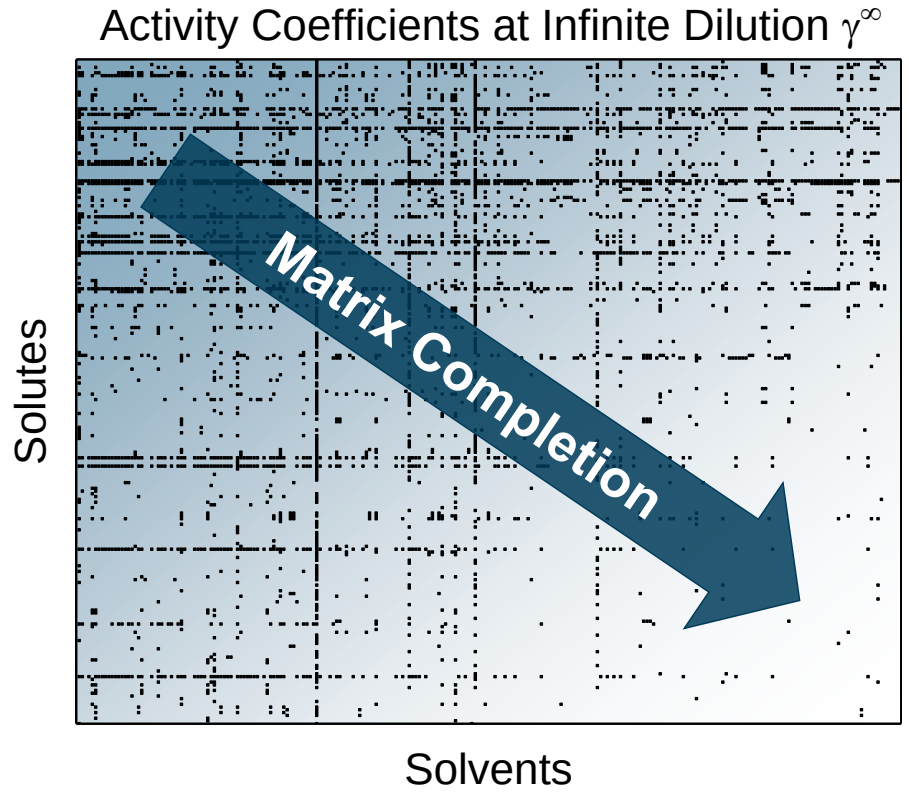

In this work, we describe a novel application of Machine Learning (ML) to the field of physical chemistry and thermodynamics: the prediction of physico-chemical properties of binary liquid mixtures by matrix completion. We focus on the prediction of a single property: the so-called activity coefficient, which is a measure of the non-ideality of a liquid mixture and of enormous relevance in practice. The interesting aspect of our approach is that no expert knowledge about the components that make up the mixture was used: all we needed was an incomplete, sparse data set of binary mixtures and their measured activity coefficients that our method was able to successfully complete. We show that this simple approach outperforms an established procedure that has been the state of the art for several decades.

ML approaches to chemical and engineering sciences date back more than 50 years ago, but the genuine exploitation of the potential of ML in these fields has only recently begun[1]. An overview of recent advances in chemical and material sciences has, e.g., been given by Ramprasad et al.[2] and Butler et al.[3] ML has already been used to predict physico-chemical properties of mixtures, including activity coefficients[4-10]. Most of these approaches are basically quantitative structure-property relationships (QSPR) methods[11] that use physical descriptors of the components as input data to characterize the considered mixtures and relate them to the property of interest by an ML algorithm, e.g., a neural network. However, the scope of these approaches is in general rather small.

Binary mixtures are of fundamental importance in chemical engineering. The properties of mixtures can in general not be described based on properties of the pure components alone. If, however, the respective properties of the binary constituent 'sub-mixtures' of a multi-component mixture are known, the properties of the multi-component mixture can often be predicted[12]. The knowledge of the properties of binary mixtures is therefore key for design and optimization of most processes in chemical engineering.

Since the experimental determination of physico-chemical properties is cumbersome, it is practically infeasible to study all binary mixtures of all relevant components. Consequently, even the largest data bases of physico-chemical properties, such as the Dortmund Data Bank (DDB)[13] and the NIST Chemistry WebBook[14], contain only information on a small fraction of the relevant mixtures. Predictive methods for physico-chemical properties are therefore needed to fill the gaps.

Predicting properties of binary liquid mixtures from first principles is not possible yet, except for simple cases. But there are phenomenological models for this, such as UNIFAC[15,16] and COSMO-RS[17], which are used for the prediction of activity coefficients. Process simulations often rely on the quality of these predictions and great effort has been taken over the last decades to parameterize these models using the available experimental data.

Activity coefficients in liquid mixtures are usually described as a function of temperature and composition; the pressure dependence is so small that it can be safely neglected in most cases. In the present study, we consider activity coefficients $\gamma_{ij}^{\infty}$ of solutes $i$ at infinite dilution in solvents $j$ at 298.15 (±1) K, which have been measured for many binary mixtures $i$ - $j$. Our basic goal is to illustrate that ML techniques are useful for predicting such properties of binary mixtures in general.

Besides $\gamma_{ij}^{\infty}$, there are many other important properties of this type, e.g. diffusion coefficients or gas solubility as described by the Henry's law constant. As data on a given property of different binary mixtures can be represented conveniently in a matrix, the appropriate ML techniques for predicting such properties are matrix completion methods (MCM). To the best of our knowledge, they have never been used before for this purpose.

The activity coefficient at infinite dilution is a key property for process design and optimization, since the concentration dependence of both activity coefficients in the binary system $i$ - $j$ can usually be predicted from $\gamma_{ij}^{\infty}$ and $\gamma_{ji}^{\infty}$. From the activity coefficients, the chemical potential of the components can be calculated, which is needed to describe chemical and phase equilibria as well as transport processes. Furthermore, as mentioned above, also activity coefficients in multi-component systems can be predicted from information on binary systems[12].

Experimental data on $\gamma_{ij}^{\infty}$ at 298.15 (±1) K are available for several thousand solute-solvent combinations. These data can be represented as the entries of a matrix, whose rows and columns correspond to the solutes $i$ and the solvents $j$, respectively. Figure 1 shows a schematic representation of the studied matrix, in which the mixtures for which experimental data are available are indicated by black squares. Filling the gaps, i.e., predicting $\gamma_{ij}^{\infty}$ for the missing entries, can be regarded as a matrix completion problem.

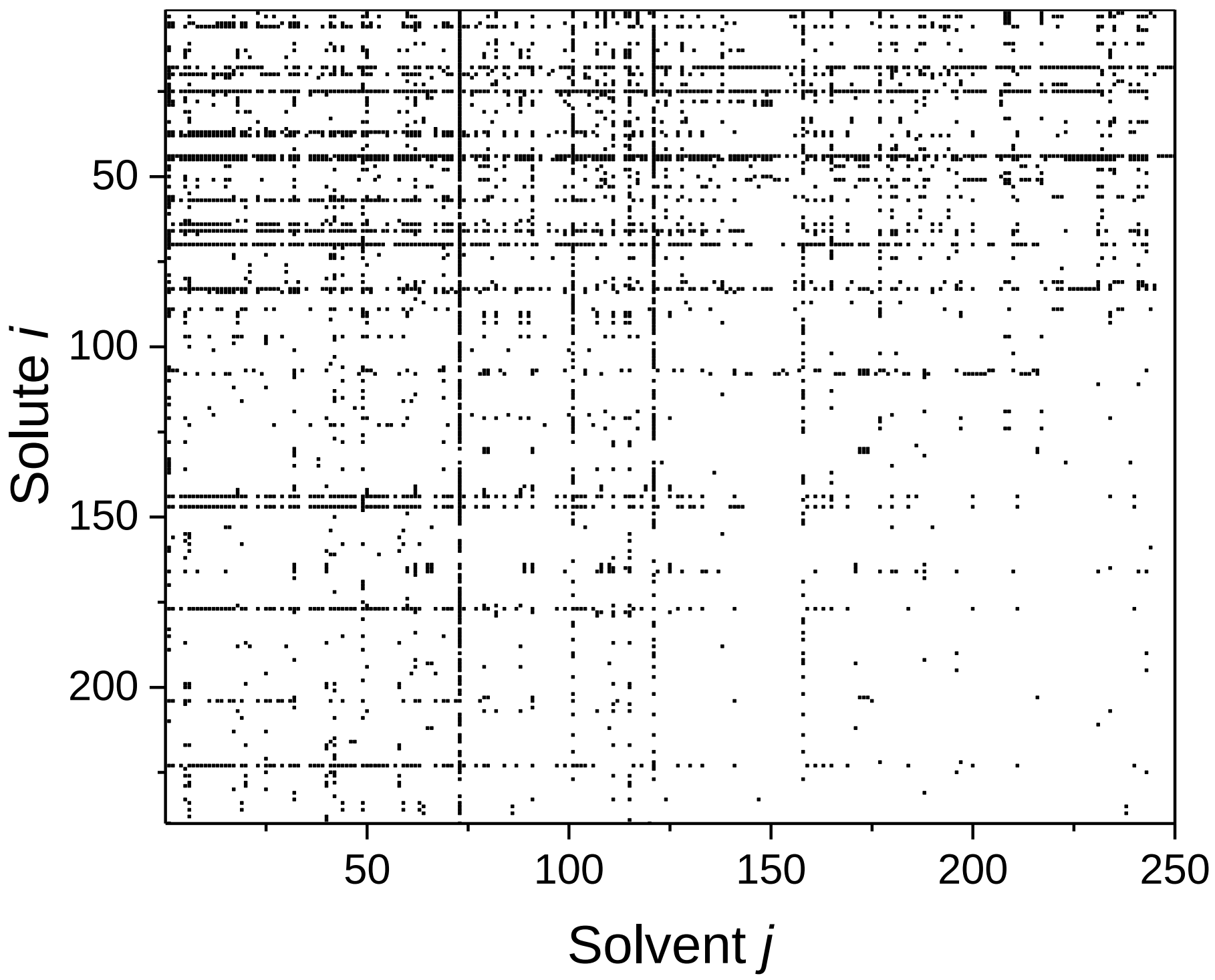

**Figure 1.** Schematic depiction of the matrix representing all possible binary mixtures of the studied 240 solutes and 250 solvents. The black squares indicate mixtures for which experimental data on the activity coefficients at infinite dilution $\gamma_{ij}^{\infty}$ at 298.15 (±1) K are available in the 2019 version of the DDB[13].

Matrix completion is well studied in ML and has become popular through the Netflix Prize[18], an open competition by Netflix that aimed at improving their recommender system for movies and TV shows. Subsequently, several matrix completion methods have been proposed and applied for various purposes[19-23].

Matrix completion problems can be addressed with different approaches. One distinguishes between content-based filtering methods[24] and collaborative filtering methods[25]. Besides the observed entries of the matrix, content-based filtering employs descriptors of the considered systems to complete the matrix. Collaborative filtering, by contrast, solely learns from the observed entries of the matrix, relying on pattern-recognition techniques to find similarities within the rows and the columns, to predict the missing entries of the partially observed matrix.

In this work, we use a collaborative filtering approach to matrix completion. Hence, we predict $\gamma_{ij}^{\infty}$ for the unobserved mixtures based only on $\gamma_{ij}^{\infty}$ of the observed mixtures, i.e., the mixtures for which experimental data are available. Furthermore, our matrix completion method follows the Bayesian approach and consists of three steps. In the first step, a generative probabilistic model of the data, i.e., $\gamma_{ij}^{\infty}$, as a function of initially unknown features of the components $i$ and $j$ is formulated. This generative model poses a probability distribution over all $\gamma_{ij}^{\infty}$ based on the component features. In the second step, the initially unknown component features are inferred by training the model to the observed $\gamma_{ij}^{\infty}$. This step is called 'inference' and requires the inversion of the generative model. Since our generative model is probabilistic, its inverse is also probabilistic and Bayesian inference yields the so-called 'posterior probability distribution', or short 'posterior', of the component features. From the posterior, among others, the most probable numbers for the features to describe the data are obtained. Since exact Bayesian inference is infeasible in nontrivial generative models, we resort to variational inference[26-28] for an efficient approximation. We use the Stan framework[29], a so-called probabilistic programming language, which automates the task of approximate Bayesian inference in a user-defined generative model. In the last step, the inferred component features are inserted in the generative model to obtain predictions for unobserved $\gamma_{ij}^{\infty}$. All modeling details, including the source code to run the Stan model, can be found in the Supporting Information. We emphasize the simplicity of the modeling framework, which can be extended in many ways.

For training the MCM, data on $\gamma_{ij}^{\infty}$ at 298.15 (±1) K for mixtures of molecular components were taken from the present version (2019) of the Dortmund Data Bank (DDB)[13]. To allow an evaluation of the proposed MCM as described below, we considered only solutes $i$ and solvents $j$ for which at least two data points, i.e., data for at least two different mixtures, are available. This results in a data set with $I = 240$ solutes and $J = 250$ solvents. These were arranged in an $I \times J$ matrix with 60000 elements, corresponding to all possible binary solute-solvent combinations, cf. Figure 1. For 4094 entries, i.e., different binary mixtures, data are available in the present version of the DDB, which corresponds to 6.8% of all elements of the matrix. The remaining 55906 entries were predicted by the MCM based on the available entries. The study was carried out using $\ln(\gamma_{ij}^{\infty})$ rather than $\gamma_{ij}^{\infty}$ for scaling purposes. Figure S1 in the Supporting Information shows the distribution of the $\ln(\gamma_{ij}^{\infty})$ values in the data set. A list of the considered solutes and solvents is given in Tables S2 and S3 in the Supporting Information, respectively.

To evaluate the predictions of the MCM, we applied leave-one-out cross-validation[30]. Therefore, the MCM was trained on all observed entries except for one. This left-out entry was then predicted by matrix completion and compared to its experimental value reported in the DDB. This procedure was repeated for all observed entries. Figure 2 shows the predictions obtained with the MCM in a parity plot over the experimental data. A histogram representation of the results is given in Figure S4 in the Supporting Information. For about 48.1% of the data, $\ln(\gamma_{ij}^{\infty})$ is predicted with an absolute error below 0.1; about 79.6% the data are predicted with an absolute error below 0.3. This performance is remarkable, especially considering that no physical descriptors of the components were used and that the experimental uncertainty of $\ln(\gamma_{ij}^{\infty})$ is typically 0.1 to 0.2.

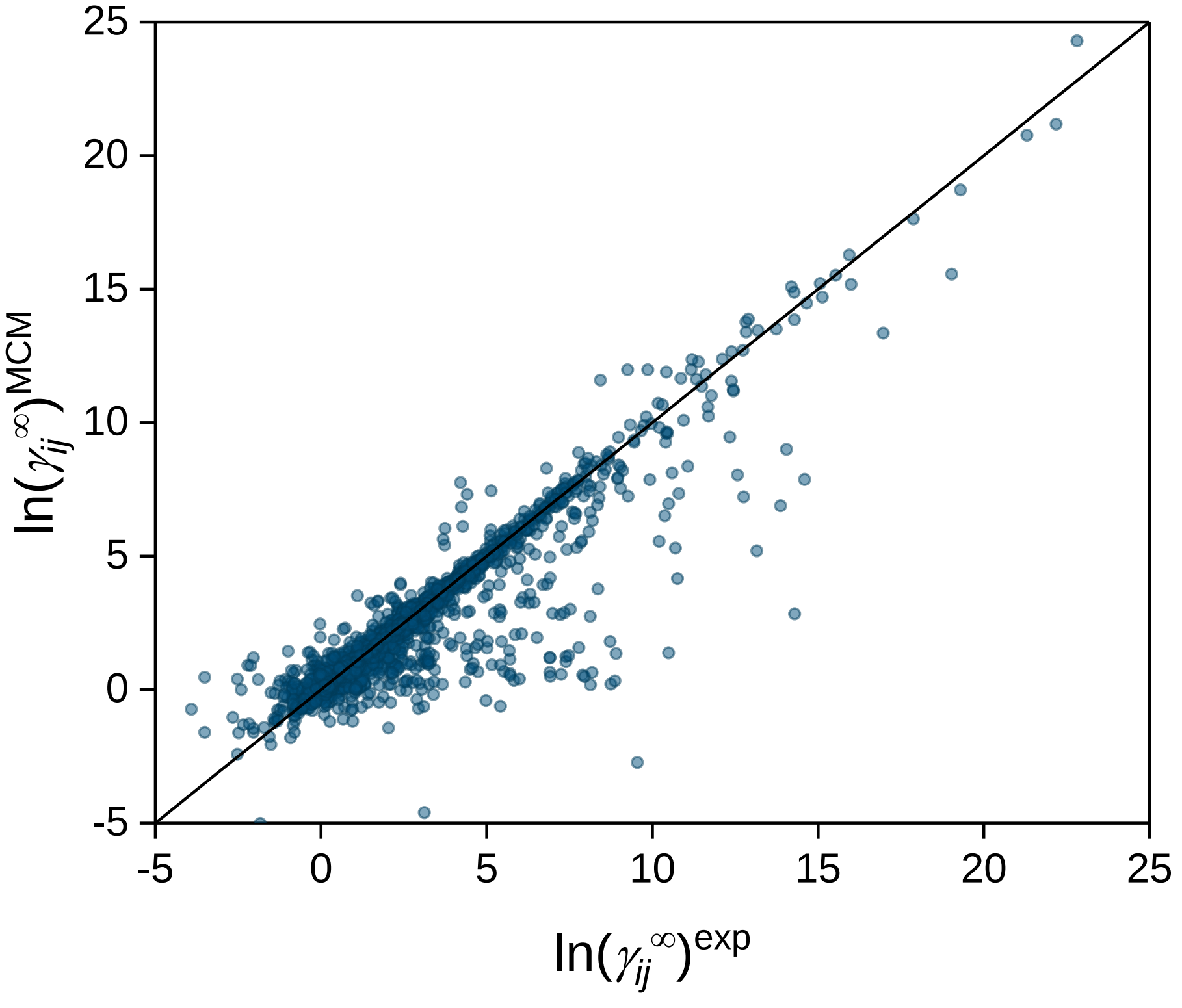

**Figure 2.** Parity plot of the predictions for $\ln(\gamma_{ij}^{\infty})$ with the proposed MCM over the corresponding experimental values (exp) from the DDB. The depicted range includes results for 99.9% of the total data set.

In the following, we compare the proposed MCM with one of the highly developed physical methods for predicting activity coefficients. Modified UNIFAC (Dortmund)[31,32], referred to simply as UNIFAC in the following, is the most successful of these methods and has been considered as the gold standard for more than 30 years. In UNIFAC, the properties of a mixture are determined by the functional groups of the molecules and their interactions. The interaction parameters are obtained by fitting them to experimental data.

With its present published parameterization, UNIFAC is able to predict the activity coefficients for 3342 of the 4094 solute-solvent combinations that are considered here. In Figure 3, we compare the predictions for this subset obtained with the proposed MCM with those from UNIFAC in a histogram. The corresponding parity plot is given in Figure S5 in the Supporting Information. The results demonstrate a better performance of the proposed MCM. As an example, the absolute error is below 0.1 for 37.4% of the predictions with UNIFAC, whereas the proposed MCM achieves the same accuracy for 50.0% of the predictions. The MCM also clearly outperforms UNIFAC in terms of mean absolute deviation and mean square error, cf. Table S1 in the Supporting Information.

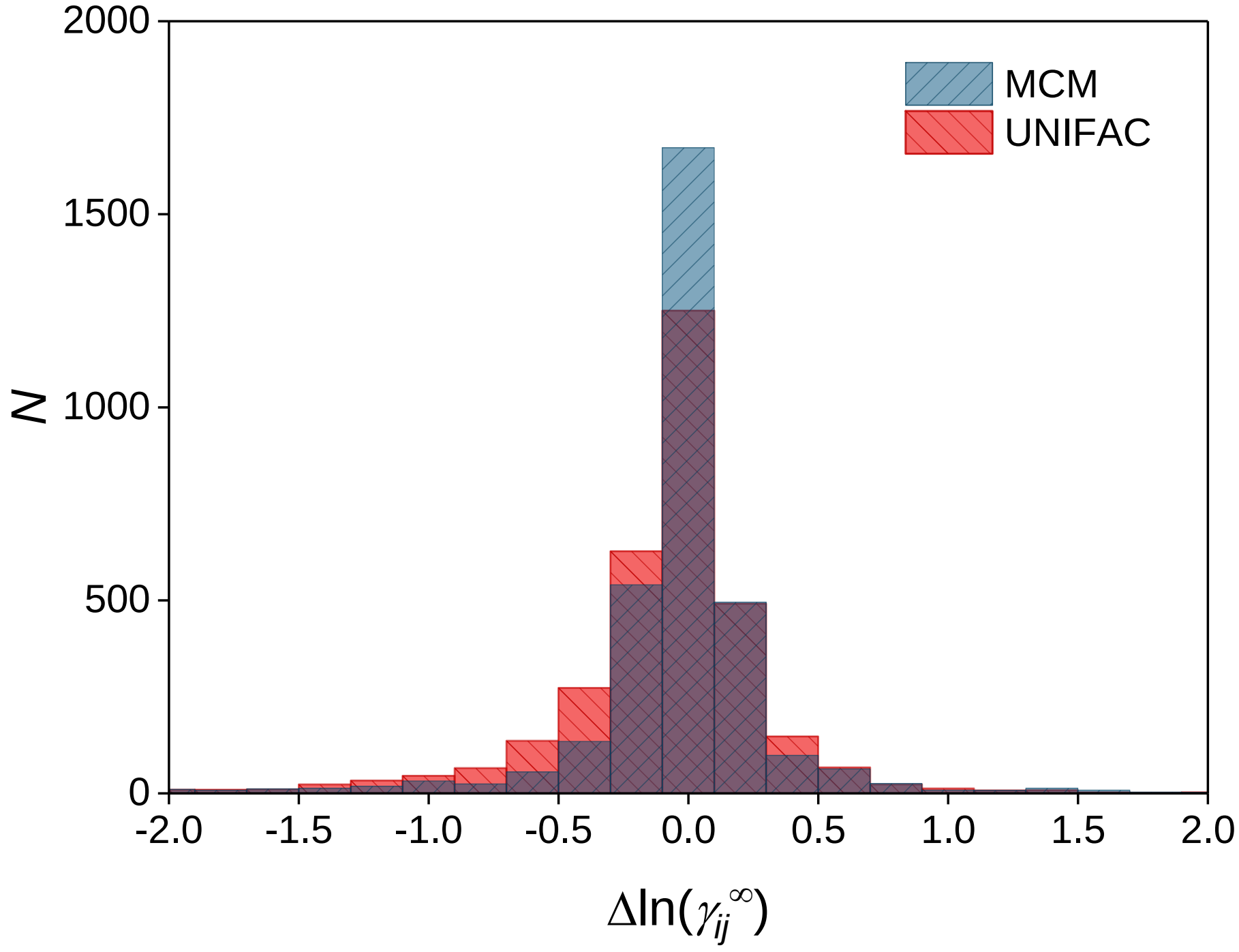

**Figure 3.** Histogram of the differences of the predictions for $\ln(\gamma_{ij}^{\infty})$ with the proposed MCM or UNIFAC and the corresponding experimental values (exp) from the DDB. $\Delta\ln(\gamma_{ij}^{\infty}) = \ln(\gamma_{ij}^{\infty})^{\text{MCM / UNIFAC}} - \ln(\gamma_{ij}^{\infty})^{\text{exp}}$. $N$ represents the number of binary mixtures $i$ - $j$ for which the differences are within the given intervals. The depicted range includes results for 96.9% of the total data set for both methods.

Besides the better performance, the proposed MCM has two additional clear advantages over UNIFAC. First, the further development of UNIFAC is extremely elaborate. UNIFAC is based on the segmentation of components into groups. Choosing these groups and determining the group parameters as well as the group interaction parameters from selected data sets is an art that is practiced by only a few specialists, several generations of which have been working on the method since it was first introduced in 1975. By contrast, matrix completion is a general concept that is easy to use, and that can be improved simply by retraining on a larger data set whenever new experimental data become available. Second, the application of UNIFAC to predict $\gamma_{ij}^{\infty}$ is limited by the availability of the required group parameters, which are elaborate to obtain as described above. For the solutes and solvents considered here, $\gamma_{ij}^{\infty}$ for less than two thirds of all binary mixtures can be predicted with UNIFAC, cf. Figure S2 in the Supporting Information. With the proposed MCM, $\gamma_{ij}^{\infty}$ for all possible combinations of the studied components can be predicted, i.e., all gaps in the matrix can be filled.

Our results demonstrate the potential of using matrix completion to predict $\gamma_{ij}^{\infty}$ in binary mixtures, but should be considered as only the first step towards using MCM for predicting physico-chemical properties of binary mixtures in general. In future work, physical descriptors will be included in the MCM algorithm. These physical descriptors could, for example, contain information on the chemical groups of the components, as they are used in UNIFAC. Further iterations could also consider other choices, such as $\sigma$-profiles of the components, as they are used in COSMO-RS[17]. It can be expected that adding such information will lead to significant improvements. A feature analysis of the MCM results could reveal structures in the data that could provide further insight to physical structure-property relations. Furthermore, the approach will be extended to other temperatures and properties. Ultimately, we conjecture that our approach only scratched the surface of what is possible and may inspire the next generation of prediction methods in chemical engineering.

**Acknowledgments**
Fabian Jirasek greatly acknowledges financial support by a postdoc fellowship of the German Academic Exchange Service (DAAD). The Fraunhofer team acknowledges funding through the Fraunhofer Cluster of Excellence »Cognitive Internet Technologies«. Stephan Mandt acknowledges funding from DARPA (HR001119S0038), NSF (FW-HTF-RM), and Qualcomm.

**Notes**
The authors declare no competing financial interests.

**Supporting Information**
Information on the used experimental data and data preprocessing. Information on the probabilistic model, variational inference, and the calculation of model predictions. Additional results for an alternative model based on a normal likelihood.

Supporting Information for

# Machine Learning in Thermodynamics: Prediction of Activity Coefficients by Matrix Completion

*Fabian Jirasek[1,2†*], Rodrigo A. S. Alves[3†], Julie Damay[4†], Robert A. Vandermeulen[3], Robert Bamler[1], Michael Bortz[4‡], Stephan Mandt[1‡], Marius Kloft[3‡], Hans Hasse[2‡]*

[1]Department of Computer Science, University of California, Irvine, USA

[2]Laboratory of Engineering Thermodynamics (LTD), TU Kaiserslautern, Kaiserslautern, Germany

[3]Machine Learning Group, Department of Computer Science, TU Kaiserslautern, Kaiserslautern, Germany

[4]Fraunhofer Institute for Industrial Mathematics ITWM, Kaiserslautern, Germany

[†]These authors contributed equally to this work.

[‡]These authors jointly directed this work.

**Corresponding Author**

*Correspondence to: fabian.jirasek@mv.uni-kl.de

## Experimental Data and Preprocessing

All data for training and evaluation of the proposed matrix completion method (MCM) were taken from the current version (2019) of the Dortmund Data Bank (DDB)[1]. All data for activity coefficients at infinite dilution $\gamma_{ij}^{\infty}$ in binary mixtures at temperatures ranging from 297.15 to 299.15 K, i.e., at 298.15 (±1) K, were adopted. The temperature dependence of activity coefficients in such narrow temperature ranges is in general small and is therefore not considered here. For several solute *i* - solvent *j* combinations, multiple results on $\gamma_{ij}^{\infty}$ in the considered temperature range are available in the DDB. For these combinations, the arithmetic mean of all available data was used for training and evaluation. The data set was further modified as follows: only molecular components were considered. Non-molecular solutes and solvents, mainly salts and ionic liquids, but also metals and components for which no molecular formula was available, were eliminated from the data set. This restriction is not mandatory, but we consider the excluded components substantially different such that it is not reasonable to model them alongside the studied components. Furthermore, to be able to evaluate the predictions of the proposed MCM by leave-one-out cross-validation, all solutes and solvents for which only data on $\gamma_{ij}^{\infty}$ in a single mixture were available were eliminated from the data set. In total, 240 solutes and 250 solvents complied with the above stated conditions and were considered in the present study. Figure S1 shows the distribution of the experimental $\gamma_{ij}^{\infty}$ values in the studied data set in a logarithmic scale.

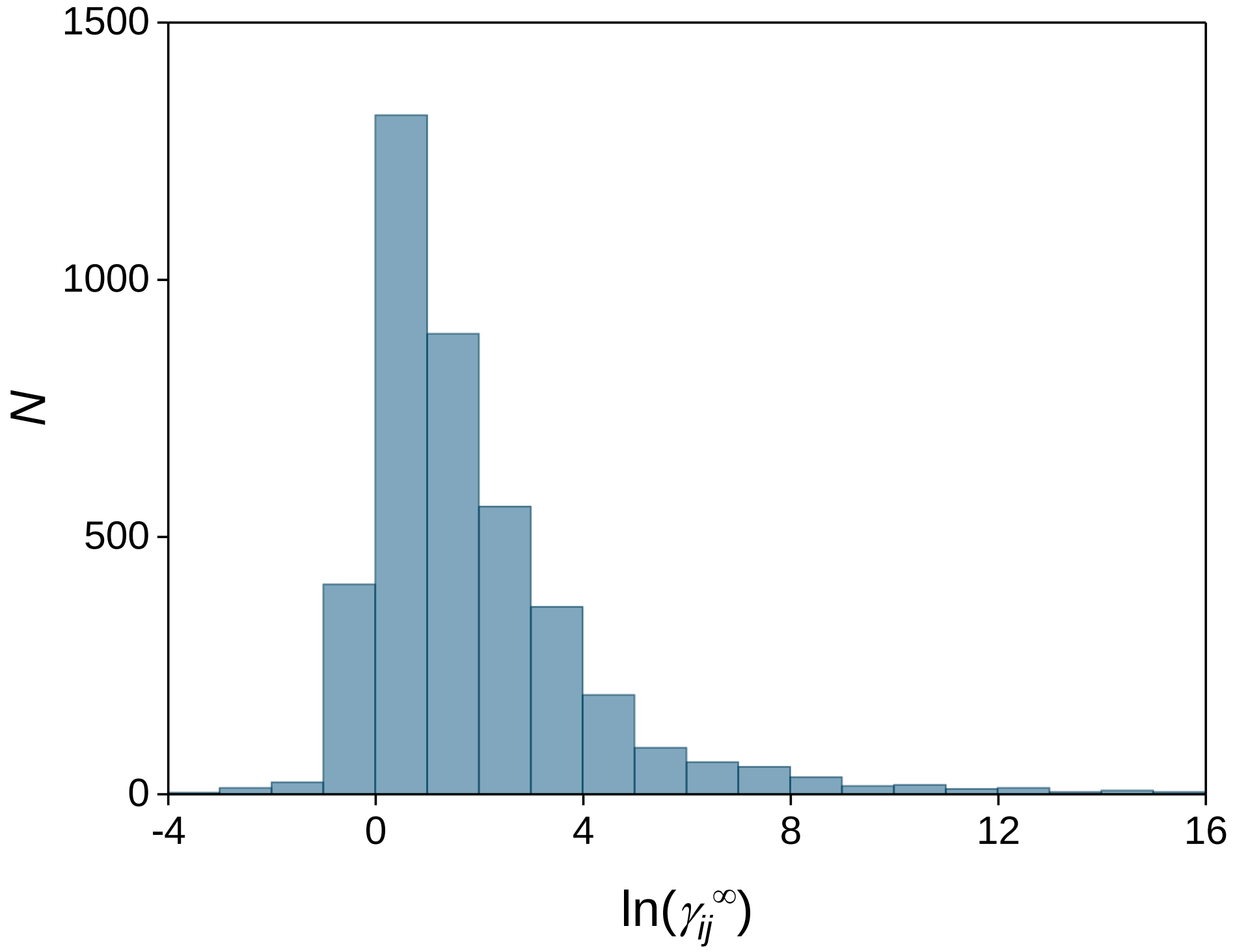

**Figure S1.** Histogram of the logarithmic values of the activity coefficients at infinite dilution $\gamma_{ij}^{\infty}$ that were used for training and testing the proposed MCM. *N* represents the number of binary mixtures *i* - *j* for which $\ln(\gamma_{ij}^{\infty})$ is within the given intervals. The depicted range includes 99.8% of the total data set.

Figure S2 shows a schematic representation of all possible binary mixtures of the studied solutes *i* and solvents *j*. The black squares indicate the mixtures for which experimental data on $\gamma_{ij}^{\infty}$ at 298.15 (±1) K are available in the present version of the DDB. Additionally, the color code indicates if the mixtures can be modeled with the present published version of modified UNIFAC (Dortmund)[2,3], simply referred to as UNIFAC in the following, or not.

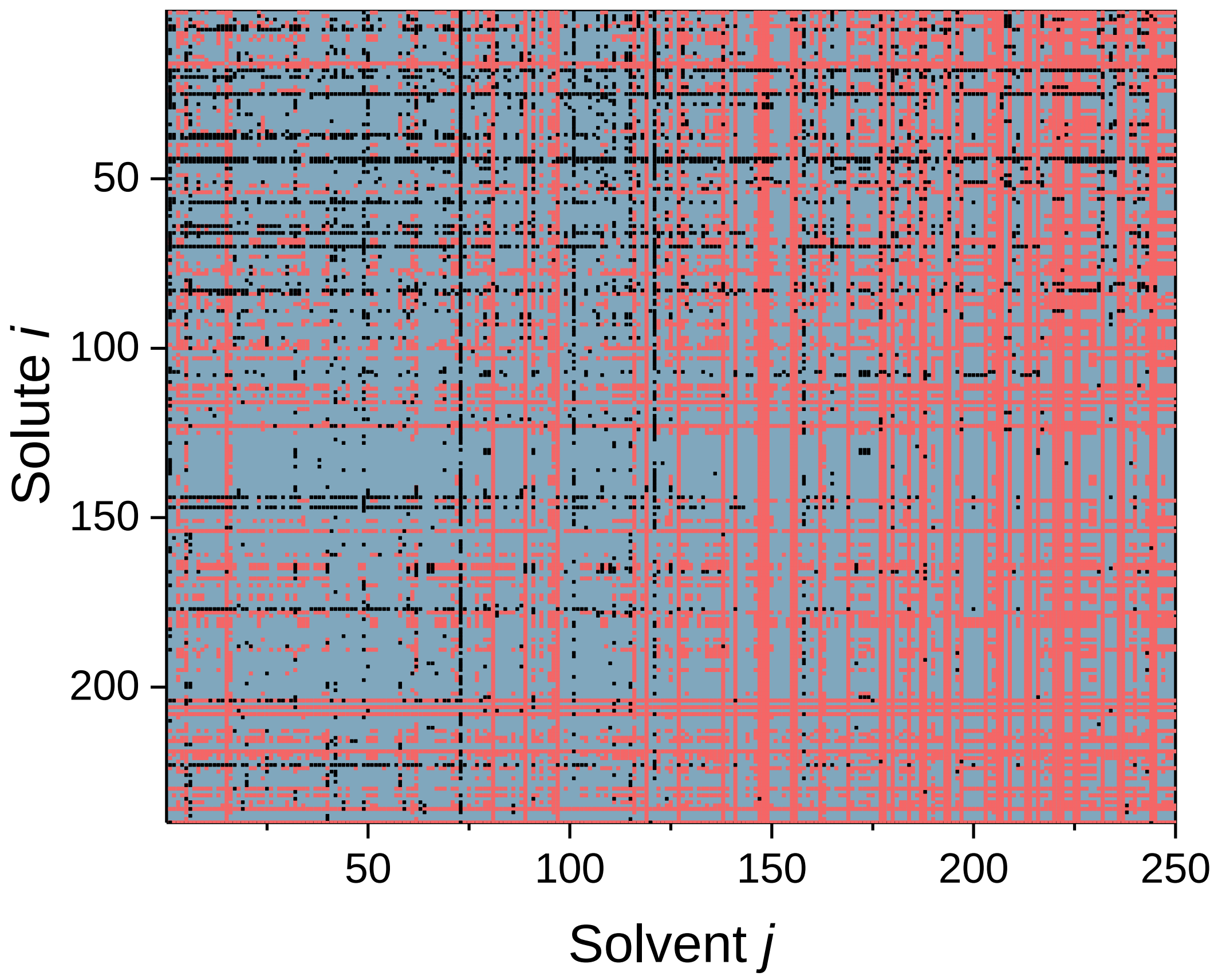

**Figure S2.** Schematic depiction of the matrix representing all possible binary mixtures of the studied 240 solutes and 250 solvents. Black: mixtures for which experimental data on $\gamma_{ij}^{\infty}$ at 298.15 (±1) K are available in the 2019 version of the DDB. Blue: mixtures for which no experimental data are available and UNIFAC can be applied. Red: mixtures for which no experimental data are available and UNIFAC with its present published parameterization cannot be applied.

Tables S2 and S3, which were moved to the end of this document for improved readability, list all studied solutes and solvents, respectively. Note that 97 components appear as both solutes and solvents. Hence, the matrix contains 97 entries that correspond to pure components. For training the MCM, the values of $\gamma_{ij}^{\infty}$ for these entries were set to 1, i.e., $\ln(\gamma_{ij}^{\infty})$ was set to 0, which follows from the definition of the activity coefficient. These entries were not considered during the evaluation.

**Probabilistic Model**

Our matrix completion method follows a Bayesian approach building on a probabilistic generative model and an inference method. The probabilistic model defines a probability distribution over all activity coefficients in logarithmic scale $\ln(\gamma_{ij}^{\infty})$ by specifying a stochastic process that generates hypothetical activity coefficients conditioned on some initially unknown, or 'latent', parameters of the components *i* and *j*. These parameters are called component features in the following. The inference method inverts the generative process and reasons about the component features for given observations, i.e., data on $\ln(\gamma_{ij}^{\infty})$.

For each solute *i* (each solvent *j*), the generative process first draws a latent feature vector $u_i$ ($v_j$) of dimension $K = 4$ from a normal distribution with zero mean and standard deviation $\sigma_0$. It then models the probability of each $\ln(\gamma_{ij}^{\infty})$ as a Cauchy distribution with scale $\lambda$ centered around the dot product of $u_i$ and $v_j$. This is called a probabilistic matrix factorization model since the large matrix of $\ln(\gamma_{ij}^{\infty})$ is modeled in terms of the product of a (smaller) tall matrix, whose rows are the solute feature vectors $u_i$, and a narrow matrix, whose columns are the solvent feature vectors $v_j$. The parameters $\sigma_0$ and $\lambda$ were set by cross-validation to $\sigma_0 = 0.8$ and $\lambda = 0.15$. Figure S3 shows our implementation of the generative model in the probabilistic programming language Stan[4], which automates the task of approximate Bayesian inference in a user-defined generative model. We also fitted a model where we replaced the Cauchy distribution by a normal distribution, see Section 'Additional Results' below.

```
1  data {
2    int<lower=0> I;          // number of solutes
3    int<lower=0> J;          // number of solvents
4    int<lower=0> K;          // number of latent dimensions
5    real ln_gamma[I,J];      // matrix of logarithmic activity coefficients
6    real<lower=0> sigma_0;   // Prior standard deviation
7    real<lower=0> lambda;    // Likelihood scale
8  }
9
10 parameters {
11   vector[K] u[I];    // solute feature vectors
12   vector[K] v[J];    // solvent feature vectors
13 }
14
15 model {
16   // prior: draw feature vectors for all solutes and solvents:
17   for (i in 1:I)
18     u[i] ~ normal(0,sigma_0);
19   for (j in 1:J)
20     v[j] ~ normal(0,sigma_0);
21
22   // likelihood: model the probability of ln_gamma as a Cauchy distribution
23   // around the dot product of the feature vectors:
24   for (i in 1:I) {
25     for (j in 1:J) {
26       if (ln_gamma[i,j] != -99) { // train to available data only
27         ln_gamma[i,j] ~ cauchy(u[i]' * v[j],lambda);
28       }
29     }
30   }
31 }
```

**Figure S3.** Stan code for the proposed matrix completion method, adapted from Kucukelbir et al.[5] Line 26 ensures that the method is only trained to the observed entries of the matrix, since all unobserved entries were set to -99 prior to the training. In an alternative model, a normal distribution was used as likelihood (line 27), cf. Section 'Additional Results' below.

## Variational Inference

The inference algorithm fits the probabilistic model to the observed data by calculating the so-called posterior probability distribution, i.e., the probability distribution over the latent feature vectors $u_i$ and $v_j$ conditioned on the observed activity coefficients. As exact posterior inference is infeasible, we resort to Gaussian mean field variational inference[5-7] (VI), which approximates the exact posterior distribution by a normal distribution for each latent feature. This process is automated by the Stan framework. In detail, VI poses a so-called variational family, i.e., a family of probability distributions over the latent feature vectors that are parameterized by so-called variational parameters, and that are considered candidates for an approximate posterior. In Gaussian mean-field VI, the variational family consists of all fully factorized normal distributions, and the variational parameters are the means and standard deviations along each coordinate of the latent space. VI then finds the element of the variational family that most closely matches the true posterior distribution by numerically minimizing the so-called Kullback-Leibler divergence from the true posterior to the approximate posterior. This can be done without having to explicitly calculate the true posterior, which would be numerically infeasible. We refer to the literature[6,7] for more background on VI.

## Calculation of Model Predictions

To predict $\ln(\gamma_{ij}^{\infty})$ for a given previously unknown solute $i$ - solvent $j$ combination, we take the means of the corresponding feature vectors $u_i$ and $v_j$ under the approximate posterior distribution that were obtained by training the model to the data. We also experimented with a variant of this method that takes the mode instead of the mean under the posterior distribution, i.e., the values for $u_i$ and $v_j$ with highest posterior probability. This so-called maximum a-posteriori (MAP) approximation is conceptionally simpler than posterior means because searching for the MAP solution can be implemented without explicitly keeping track of uncertainties. However, we found posterior means to be more robust to outliers in the data set than MAP. Improved robustness compared to MAP is a known property of VI[8]. When we report predictions for $\ln(\gamma_{ij}^{\infty})$ in this work, the prediction is always based on a model where the solute $i$ - solvent $j$ combination that we predict was excluded from the observed data in the inference process. This ensures that the method cannot cheat by predicting the value of $\ln(\gamma_{ij}^{\infty})$ from the training data.

## Additional Results

Figure S4 shows a histogram of the differences of the predictions for $\ln(\gamma_{ij}^{\infty})$ with the proposed MCM and the corresponding experimental values from the DDB for the complete data set. Figure S4 is an alternative representation of the results shown in Figure 2 in the manuscript.

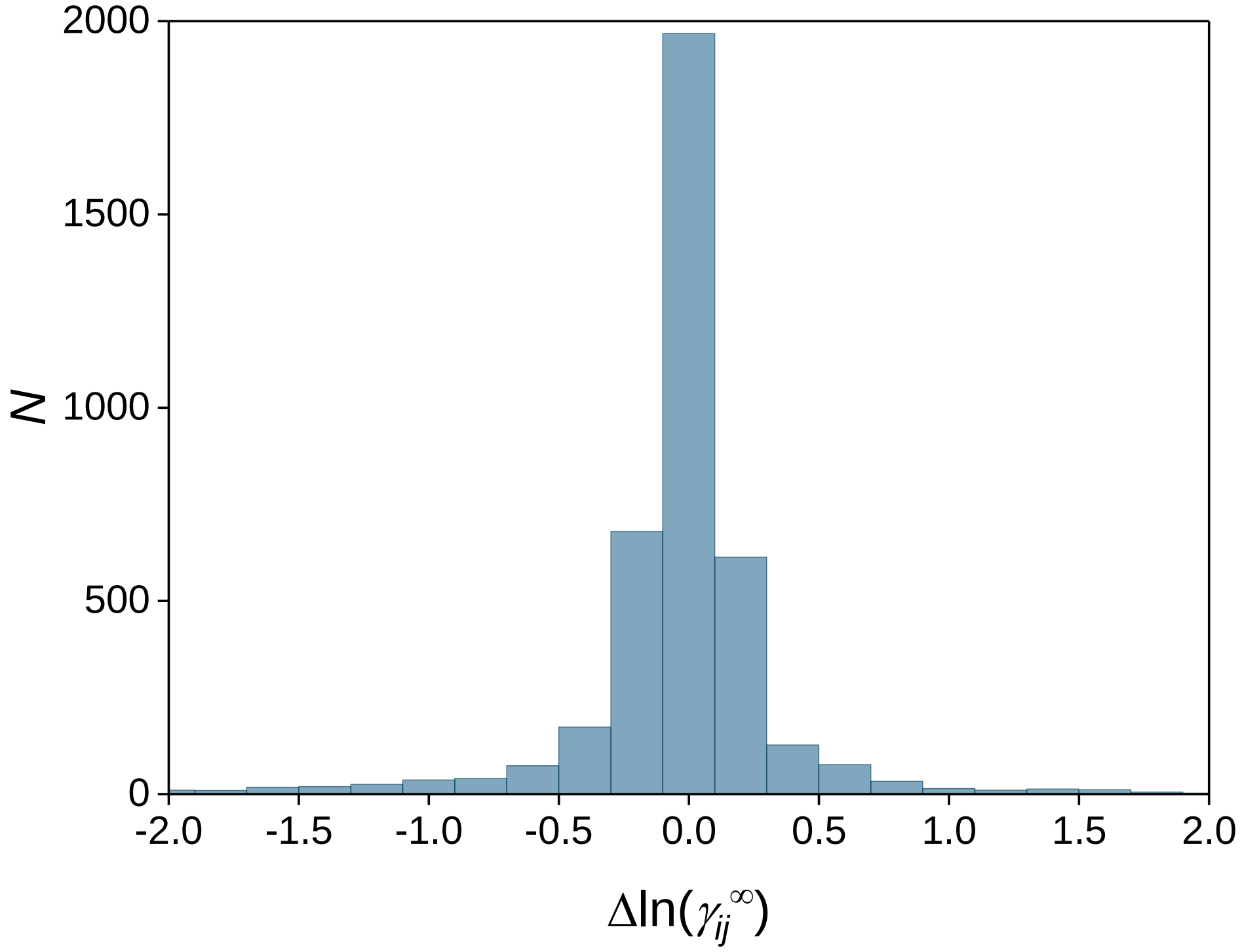

**Figure S4.** Histogram of the differences of the predictions for $\ln(\gamma_{ij}^{\infty})$ with the proposed MCM and the corresponding experimental values (exp) from the DDB: $\Delta\ln(\gamma_{ij}^{\infty}) = \ln(\gamma_{ij}^{\infty})^{\mathrm{MCM}} - \ln(\gamma_{ij}^{\infty})^{\mathrm{exp}}$. $N$ represents the number of binary mixtures $i$ - $j$ for which the differences are within the given intervals. The depicted range includes results for 96.6% of the total data set.

Figure S5 shows a parity plot of the predictions for $\ln(\gamma_{ij}^{\infty})$ with the proposed MCM and UNIFAC over the corresponding experimental values from the DDB. Only predictions for mixtures that can be modeled with UNIFAC are shown for both methods. Figure S5 is an alternative representation of the results shown in Figure 3 in the manuscript.

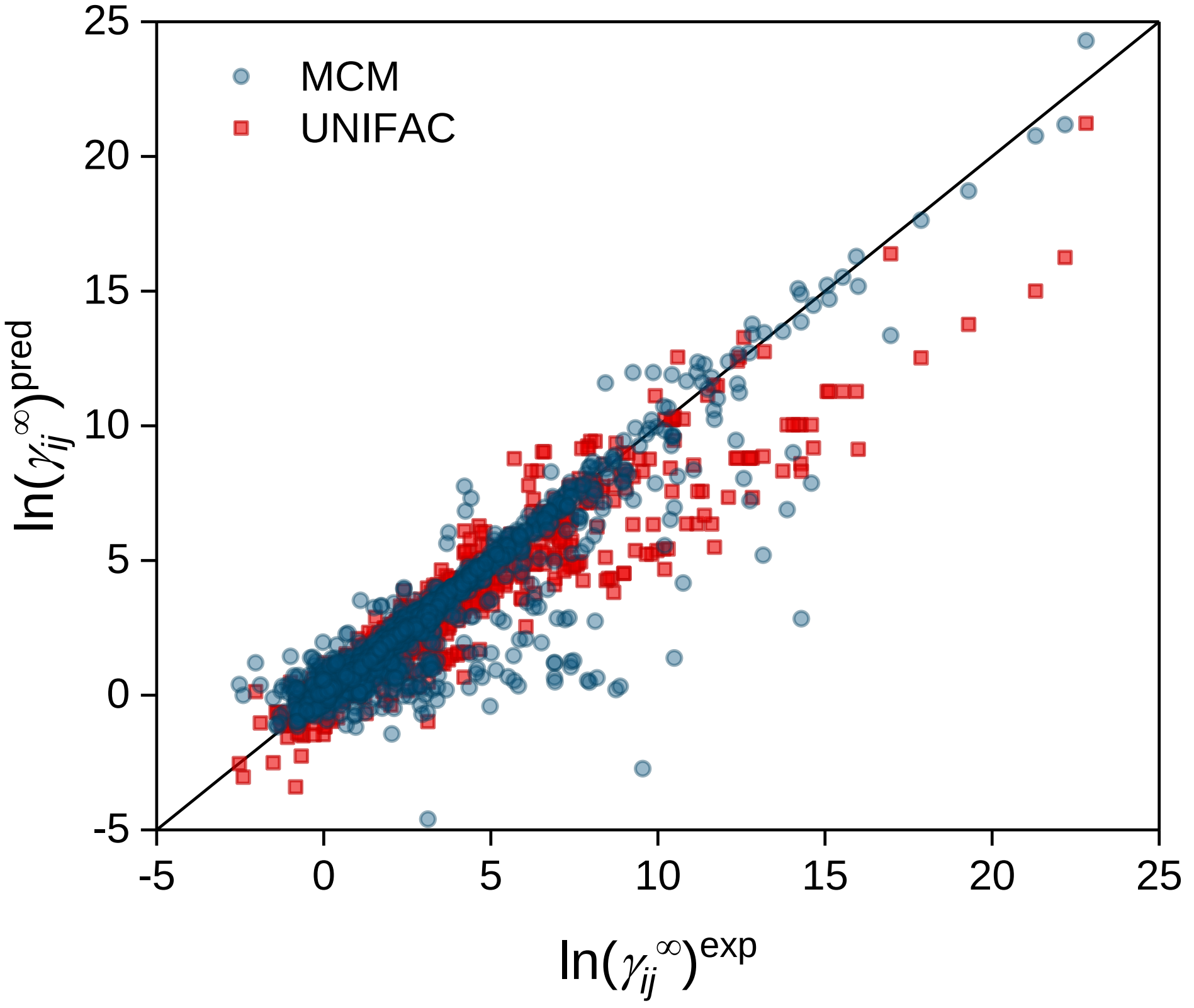

**Figure S5.** Parity plot of the predictions (pred) for $\ln(\gamma_{ij}^{\infty})$ with the proposed MCM and UNIFAC over the corresponding experimental values (exp) from the DDB. The depicted range includes results for 99.9% (MCM) and 99.7% (UNIFAC) of the data set.

In the following, predictions from the alternative model that uses a normal distribution instead of a Cauchy distribution as likelihood, cf. previous section, are shown for the same data sets as in the manuscript. The presentation of the results is essentially the same as in Figures S4 and S5 and in Figures 2 and 3 in the manuscript. The predictive power of both MCMs is similar and both outperform the state-of-the-art physical method UNIFAC. This can also be seen by considering the mean absolute deviation (MAD) and the mean square error (MSE) of the predictions compared to the experimental data, cf. Table S1.

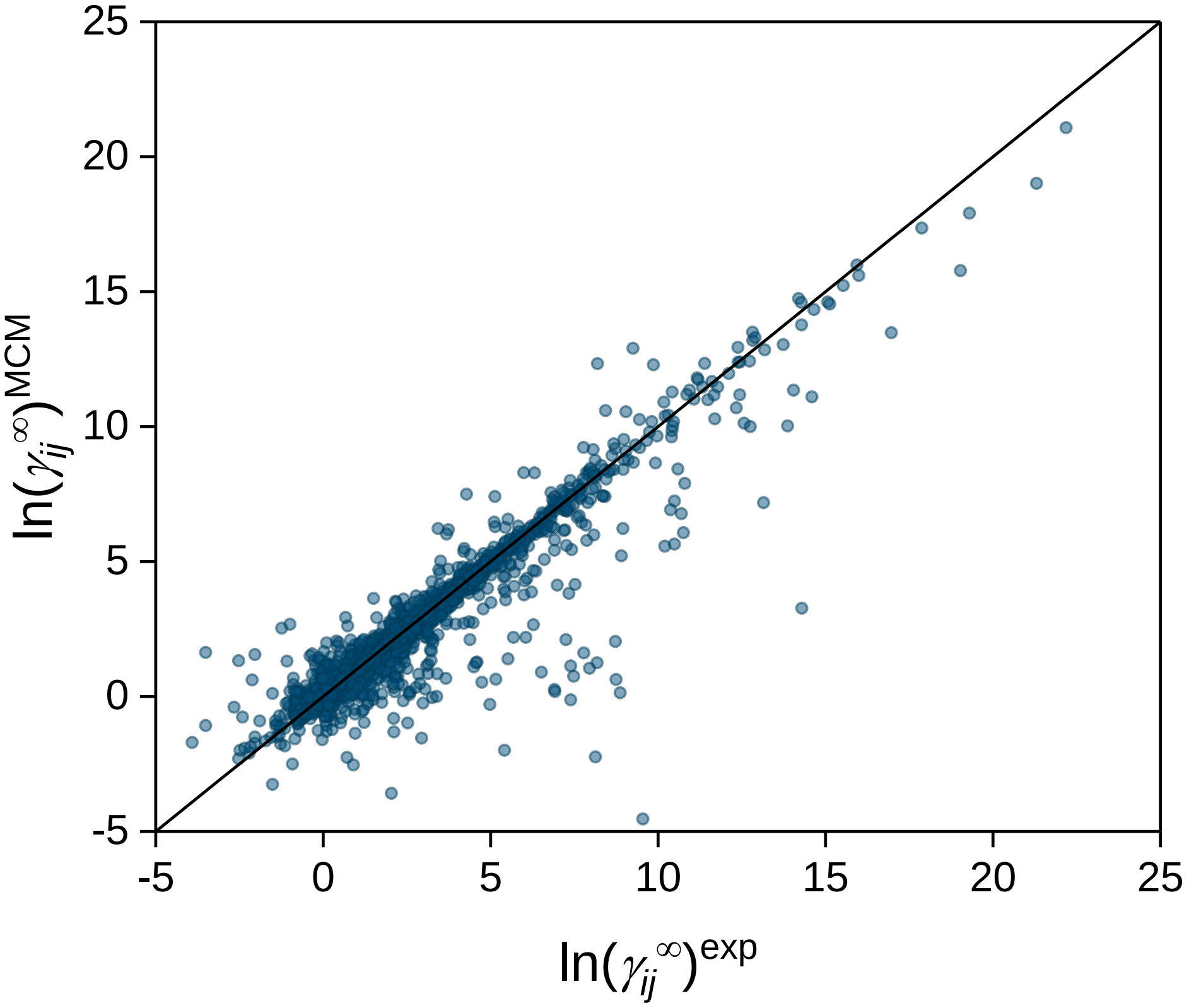

**Figure S6.** Parity plot of the predictions for $\ln(\gamma_{ij}^{\infty})$ with the alternative MCM over the corresponding experimental values (exp) from the DDB. The depicted range includes results for 99.9% of the total data set.

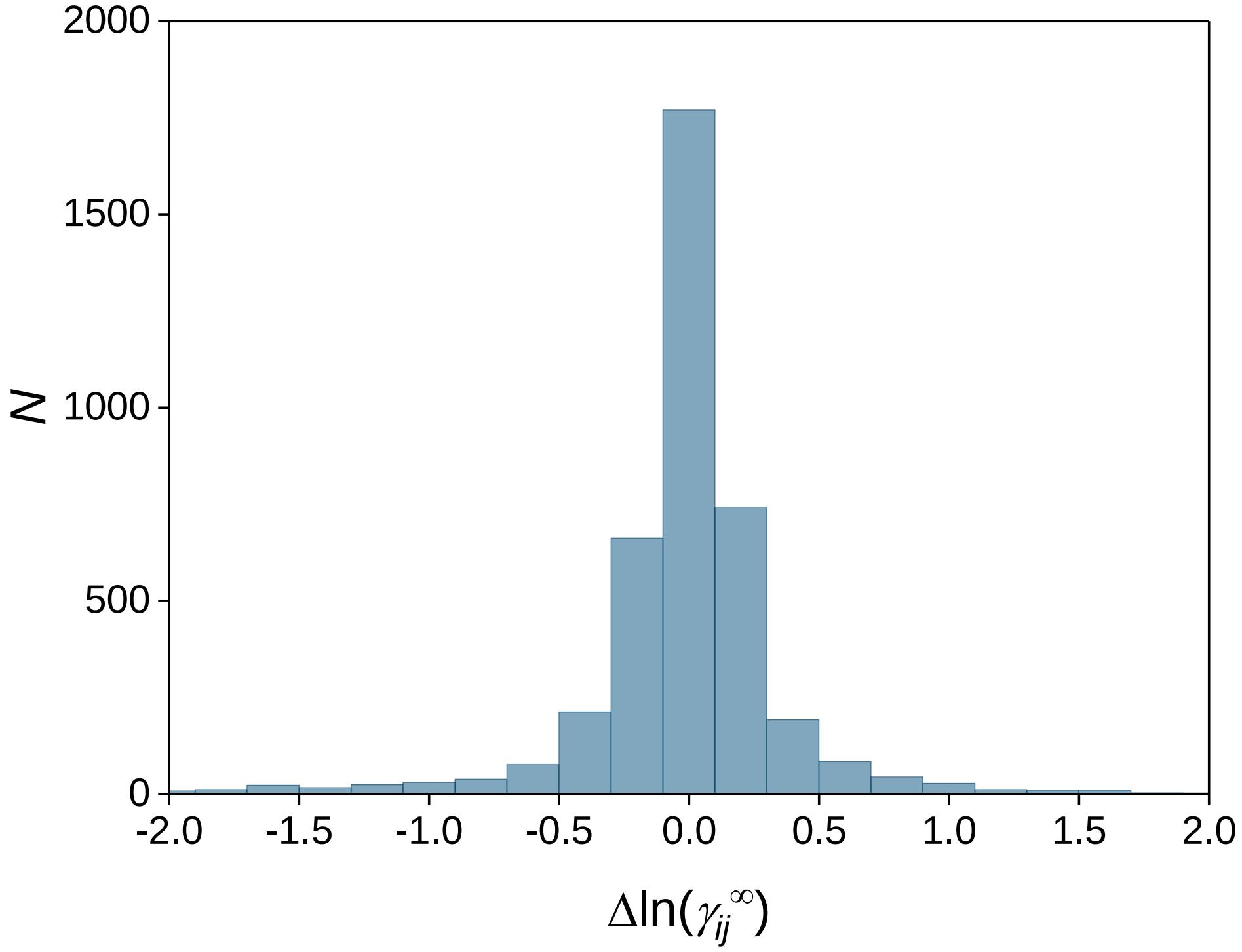

**Figure S7.** Histogram of the differences of the predictions for $\ln(\gamma_{ij}^{\infty})$ with the alternative MCM and the corresponding experimental values (exp) from the DDB: $\Delta\ln(\gamma_{ij}^{\infty}) = \ln(\gamma_{ij}^{\infty})^{\mathrm{MCM}} - \ln(\gamma_{ij}^{\infty})^{\mathrm{exp}}$. $N$ represents the number of binary mixtures $i$ - $j$ for which the differences are within the given intervals. The depicted range includes results for 97.7% of the total data set.

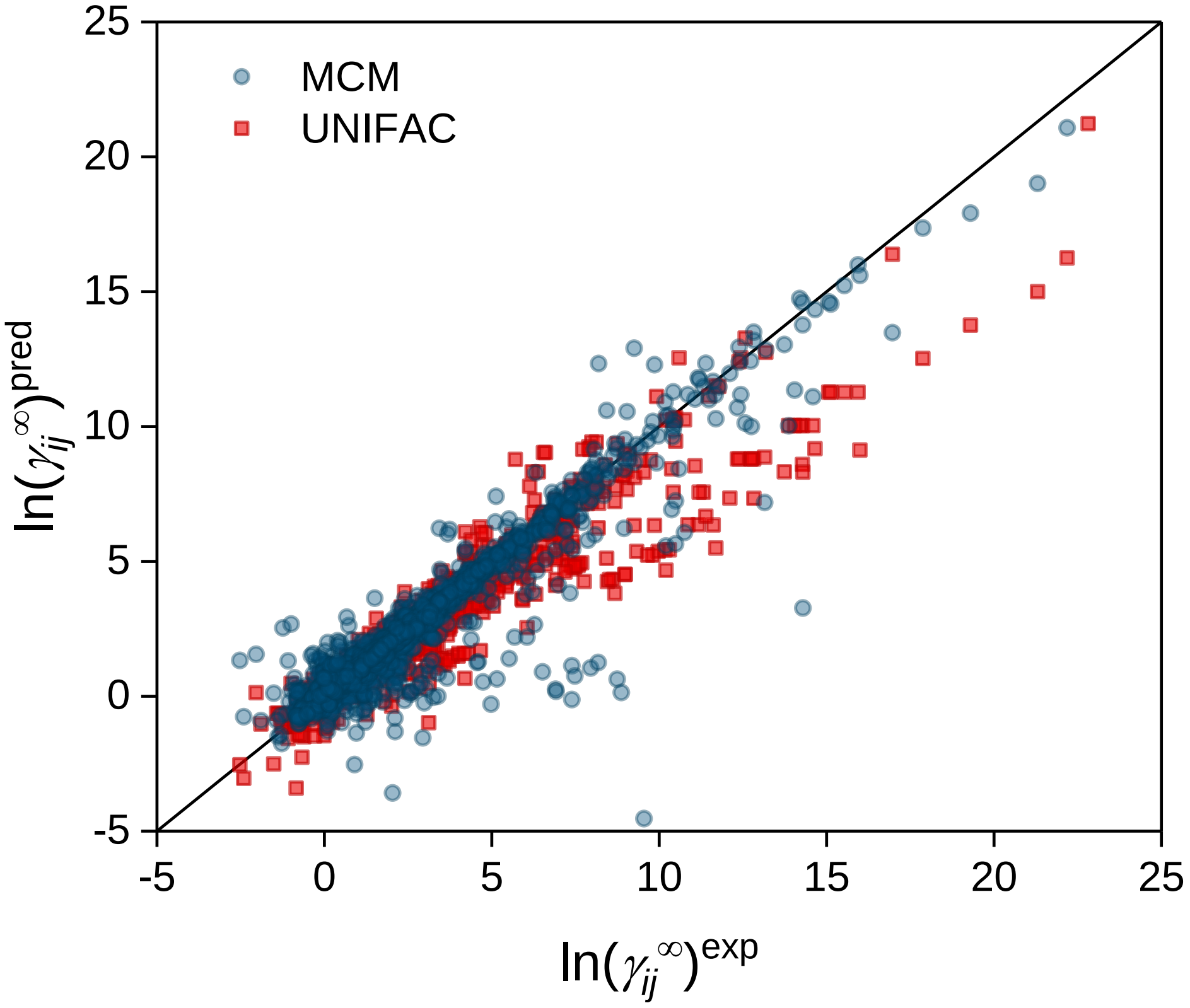

**Figure S8.** Parity plot of the predictions (pred) for $\ln(\gamma_{ij}^{\infty})$ with the alternative MCM and UNIFAC over the corresponding experimental values (exp) from the DDB. Only results for mixtures that can be modeled with UNIFAC are shown. The depicted range includes results for 99.9% (MCM) and 99.7% (UNIFAC) of the data set.

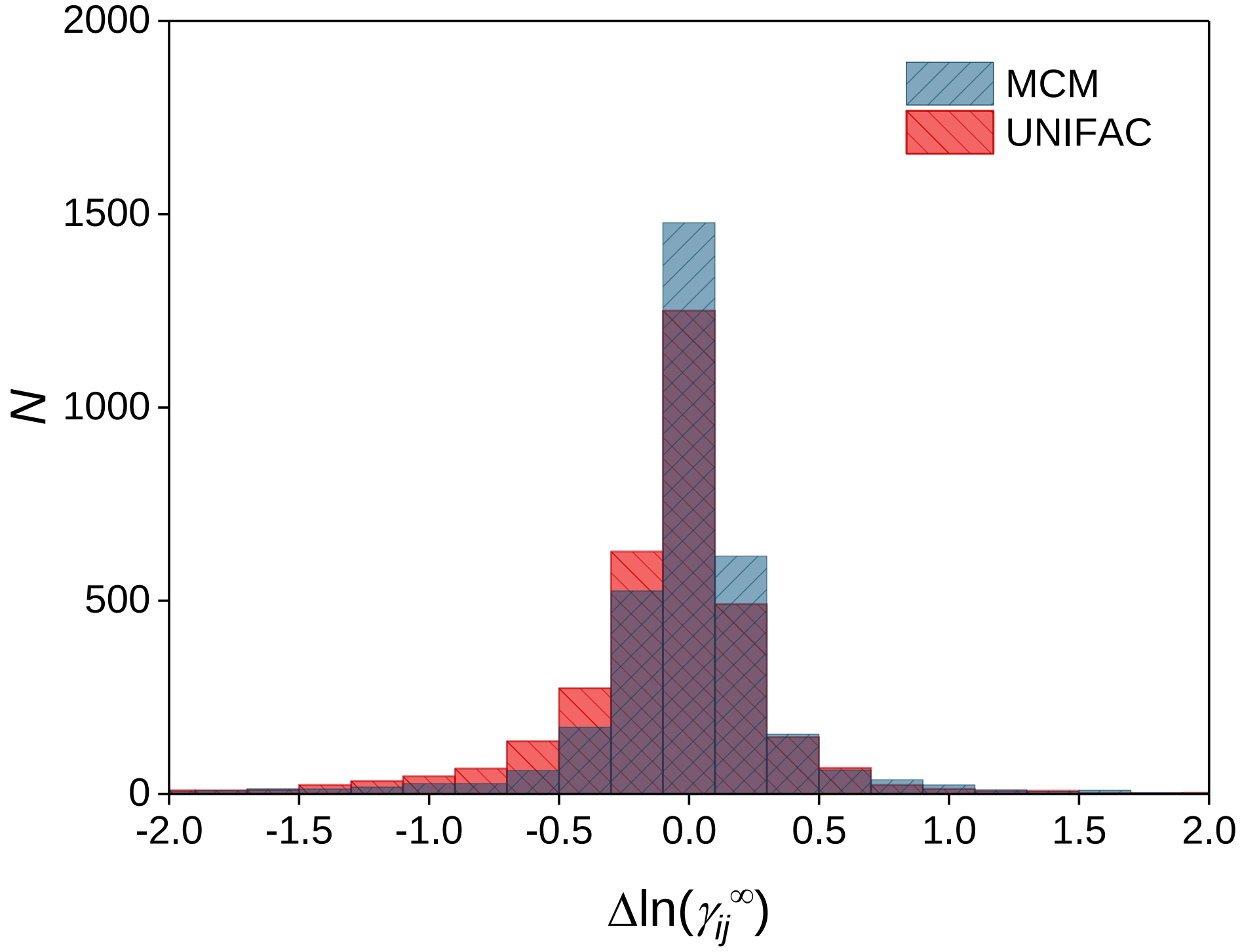

**Figure S9.** Histogram of the differences of the predictions for $\ln(\gamma_{ij}^{\infty})$ with the alternative MCM or UNIFAC and the corresponding experimental values (exp) from the DDB: $\Delta\ln(\gamma_{ij}^{\infty}) = \ln(\gamma_{ij}^{\infty})^{\text{MCM / UNIFAC}} - \ln(\gamma_{ij}^{\infty})^{\text{exp}}$. $N$ represents the number of binary mixtures $i$ - $j$ for which the differences are within the given intervals. Only results for mixtures that can be modeled with UNIFAC are shown. The depicted range includes results for 97.6% of the data set for the proposed MCM and 96.9% for UNIFAC.

**Table S1.** Mean absolute deviation (MAD) and mean square error (MSE) of the predictions with the proposed MCMs and UNIFAC referred to the experimental data in all cases. 'Cauchy' and 'Normal' refer to the likelihood of the respective methods. Two data sets were considered: the complete data set, cf. Figures 2 (in the manuscript), S4, S6, and S7, and a smaller data set containing only mixtures for which UNIFAC yields predictions, cf. Figures 3 (in the manuscript), S5, S8, and S9.

| | **Complete data set** | | **Data selection** | |
|---|---|---|---|---|
| **Method** | **MAD** | **MSE** | **MAD** | **MSE** |
| **MCM 'Cauchy'** | 0.336 | 0.825 | 0.316 | 0.773 |
| **MCM 'Normal'** | 0.315 | 0.667 | 0.305 | 0.643 |
| **UNIFAC** | n.a. | n.a. | 0.635 | 36.638 |

**Table S2.** Overview of the components that were considered as solutes in the present work. All information is adopted from the Dortmund Data Bank (DDB)[1]. In the last column, the group split according to modified UNIFAC (Dortmund)[2,3] is given, if applicable: the last three digits of each number define the subgroup, whereas with the first (two) digit(s) the count of the respective group per molecule is given.

| Component name | Chemical formula | CAS number | UNIFAC groups |
|---|---|---|---|
| Acetaldehyde | C2H4O | 75-07-0 | 1001, 1020 |
| Acetonitrile | C2H3N | 75-05-8 | 1040 |
| Acetone | C3H6O | 67-64-1 | 1001, 1018 |
| Ethyl bromide | C2H5Br | 74-96-4 | 1001, 1002, 1064 |
| Ethyl iodide | C2H5I | 75-03-6 | 1001, 1002, 1063 |
| Ethanol | C2H6O | 64-17-5 | 1001, 1002, 1014 |
| Diethyl ether | C4H10O | 60-29-7 | 2001, 1002, 1025 |
| Formic acid ethyl ester | C3H6O2 | 109-94-4 | 1023, 1001, 1002 |
| Aniline | C6H7N | 62-53-3 | 5009, 1036 |
| Methoxybenzene | C7H8O | 100-66-3 | 5009, 1010, 1024 |
| Ethyl acetate | C4H8O2 | 141-78-6 | 1001, 1002, 1021 |
| 2-Butanol | C4H10O | 78-92-2 | 2001, 1002, 1003, 1014 |
| Ethylbenzene | C8H10 | 100-41-4 | 1001, 5009, 1012 |
| Bromobenzene | C6H5Br | 108-86-1 | 5009, 1010, 1064 |
| Chlorobenzene | C6H5Cl | 108-90-7 | 5009, 1053 |
| Benzonitrile | C7H5N | 100-47-0 | n.a. |
| Nitrobenzene | C6H5NO2 | 98-95-3 | 5009, 1057 |
| Benzene | C6H6 | 71-43-2 | 6009 |
| 1-Butanol | C4H10O | 71-36-3 | 1001, 3002, 1014 |
| 2-Butanone | C4H8O | 78-93-3 | 1001, 1002, 1018 |
| n-Butane | C4H10 | 106-97-8 | 2001, 2002 |
| Butyl chloride | C4H9Cl | 109-69-3 | 1001, 2002, 1044 |
| Chloroform | CHCl3 | 67-66-3 | 1050 |
| 3-Methylphenol | C7H8O | 108-39-4 | 4009, 1011, 1017 |
| Cyclohexane | C6H12 | 110-82-7 | 6002 |
| Cyclopentane | C5H10 | 287-92-3 | 5002 |
| Cyclohexene | C6H10 | 110-83-8 | 4002, 1006 |
| Methylcyclohexane | C7H14 | 108-87-2 | 1001, 5002, 1003 |
| Methylcyclopentane | C6H12 | 96-37-7 | 1001, 4002, 1003 |
| Dibutyl ether | C8H18O | 142-96-1 | 2001, 5002, 1025 |
| Decane | C10H22 | 124-18-5 | 2001, 8002 |
| 1,1-Dichloroethane | C2H4Cl2 | 75-34-3 | 1001, 1048 |
| 1,2-Dichloroethane | C2H4Cl2 | 107-06-2 | 2044 |
| Dichloromethane | CH2Cl2 | 75-09-2 | 1047 |
| 1,2-Dichloropropane | C3H6Cl2 | 78-87-5 | 1001, 1044, 1045 |
| N,N-Dimethylformamide | C3H7NO | 68-12-2 | 1072 |
| 2,4-Dimethylpentane | C7H16 | 108-08-7 | 4001, 1002, 2003 |
| 1,4-Dioxane | C4H8O2 | 123-91-1 | 2002, 2027 |
| Dodecane | C12H26 | 112-40-3 | 2001, 10002 |
| Benzaldehyde | C7H6O | 100-52-7 | 5009, 1010, 1020 |
| Butyl acetate | C6H12O2 | 123-86-4 | 1001, 3002, 1021 |
| Methyl acetate | C3H6O2 | 79-20-9 | 1001, 1021 |
| Acetic acid | C2H4O2 | 64-19-7 | 1001, 1042 |
| Hexane | C6H14 | 110-54-3 | 2001, 4002 |
| Heptane | C7H16 | 142-82-5 | 2001, 5002 |
| 2-Heptanone | C7H14O | 110-43-0 | 1001, 4002, 1018 |
| 2-Methylbutane | C5H12 | 78-78-4 | 3001, 1002, 1003 |
| 2-Propanol | C3H8O | 67-63-0 | 2001, 1003, 1014 |
| Diisopropyl ether | C6H14O | 108-20-3 | 4001, 1003, 1026 |
| 2,2,4-Trimethylpentane | C8H18 | 540-84-1 | 5001, 1002, 1003, 1004 |
| Isoprene | C5H8 | 78-79-5 | 1001, 1005, 1007 |
| Methyl iodide | CH3I | 74-88-4 | 1001, 1063 |
| 1-Hexene | C6H12 | 592-41-6 | 1001, 3002, 1005 |
| Hexylamine | C6H15N | 111-26-2 | 1001, 4002, 1029 |
| 1-Methylnaphthalene | C11H10 | 90-12-0 | 7009, 1011, 2010 |
| Methanol | CH4O | 67-56-1 | 1015 |
| 2-Methylpentane | C6H14 | 107-83-5 | 3001, 2002, 1003 |
| 3-Methylpentane | C6H14 | 96-14-0 | 3001, 2002, 1003 |
| Butylbenzene | C10H14 | 104-51-8 | 1001, 2002, 5009, 1012 |
| 4-Methyl-2-pentanone | C6H12O | 108-10-1 | 2001, 1002, 1003, 1018 |
| 4-Methylpyridine | C6H7N | 108-89-4 | 1001, 1038 |
| 2-Methyl-1-propanol | C4H10O | 78-83-1 | 2001, 1002, 1003, 1014 |
| Naphthalene | C10H8 | 91-20-3 | 8009, 2010 |
| Nitromethane | CH3NO2 | 75-52-5 | 1054 |
| 1-Nitropropane | C3H7NO2 | 108-03-2 | 1001, 1002, 1055 |
| Octane | C8H18 | 111-65-9 | 2001, 6002 |
| 1-Octene | C8H16 | 111-66-0 | 1001, 5002, 1005 |
| 2-Methylphenol | C7H8O | 95-48-7 | 4009, 1011, 1017 |
| 4-Methylphenol | C7H8O | 106-44-5 | 4009, 1011, 1017 |
| Pentane | C5H12 | 109-66-0 | 2001, 3002 |
| 1-Pentanol | C5H12O | 71-41-0 | 1001, 4002, 1014 |
| 2-Pentanone | C5H10O | 107-87-9 | 1001, 2002, 1018 |
| Phenol | C6H6O | 108-95-2 | 5009, 1017 |
| 1-Propanol | C3H8O | 71-23-8 | 1001, 2002, 1014 |
| Propionic acid | C3H6O2 | 79-09-4 | 1001, 1002, 1042 |
| Pyridine | C5H5N | 110-86-1 | 1037 |
| Carbon disulfide | CS2 | 75-15-0 | 1058 |
| Dimethyl sulfoxide | C2H6OS | 67-68-5 | 1067 |
| tert-Butanol | C4H10O | 75-65-0 | 3001, 1004, 1014 |
| 1,2,3,4-Tetrahydronaphthalene | C10H12 | 119-64-2 | 2002, 4009, 2012 |
| Tetrachloromethane | CCl4 | 56-23-5 | 1052 |
| Tetrahydrofuran | C4H8O | 109-99-9 | 3002, 1027 |
| Toluene | C7H8 | 108-88-3 | 5009, 1011 |
| Triethylamine | C6H15N | 121-44-8 | 3001, 2002, 1035 |
| 1,1,2-Trichloroethane | C2H3Cl3 | 79-00-5 | 1044, 1048 |
| Tetrachloroethylene | C2Cl4 | 127-18-4 | 1070, 4069 |
| 1,1,1-Trichloroethane | C2H3Cl3 | 71-55-6 | 1001, 1051 |
| Trichloroethylene | C2HCl3 | 79-01-6 | 1008, 3069 |
| Water | H2O | 7732-18-5 | 1016 |
| m-Xylene | C8H10 | 108-38-3 | 4009, 2011 |
| p-Xylene | C8H10 | 106-42-3 | 4009, 2011 |
| Nitroethane | C2H5NO2 | 79-24-3 | 1001, 1055 |
| Fluorobenzene | C6H5F | 462-06-6 | 5009, 1071 |
| 1,1,2,2-Tetrachloroethane | C2H2Cl4 | 79-34-5 | 2048 |
| Propanoic acid ethyl ester | C5H10O2 | 105-37-3 | 2001, 1002, 1022 |
| Isoamyl acetate | C7H14O2 | 123-92-2 | 2001, 2002, 1003, 1021 |
| tert-Butyl chloride | C4H9Cl | 507-20-0 | 3001, 1046 |
| N-Methylformamide | C2H5NO | 123-39-7 | n.a. |
| N,N-Dimethylacetamide | C4H9NO | 127-19-5 | 1001, 1097 |
| Acrylonitrile | C3H3N | 107-13-1 | 1068 |
| Propane | C3H8 | 74-98-6 | 2001, 1002 |
| Propyl acetate | C5H10O2 | 109-60-4 | 1001, 2002, 1021 |
| Butylamine | C4H11N | 109-73-9 | 1001, 2002, 1029 |
| Cyclopentanone | C5H8O | 120-92-3 | 3002, 1019 |
| Cyclohexanone | C6H10O | 108-94-1 | 4002, 1019 |
| Cyclohexanol | C6H12O | 108-93-0 | 5002, 1003, 1014 |
| 1-Pentene | C5H10 | 109-67-1 | 1001, 2002, 1005 |
| 2-Methyl-2-butene | C5H10 | 513-35-9 | 3001, 1008 |
| 2-Methyl-1-butene | C5H10 | 563-46-2 | 2001, 1002, 1007 |
| 3-Methyl-1-butanol | C5H12O | 123-51-3 | 2001, 2002, 1003, 1014 |
| Thiophene | C4H4S | 110-02-1 | 1106 |
| N-Methyl-2-pyrrolidone | C5H9NO | 872-50-4 | 1085 |
| 3-Pentanone | C5H10O | 96-22-0 | 2001, 1002, 1019 |
| Methyl formate | C2H4O2 | 107-31-3 | 1023, 1001 |
| 1-Hexanol | C6H14O | 111-27-3 | 1001, 5002, 1014 |
| Perfluoro-n-heptane | C7F16 | 335-57-9 | 2074, 5075 |
| 2,3-Dimethylpentane | C7H16 | 565-59-3 | 4001, 1002, 2003 |
| Butyraldehyde | C4H8O | 123-72-8 | 1001, 2002, 1020 |
| 1,3-Cyclopentadiene | C5H6 | 542-92-7 | 1002, 2006 |
| 2-Methylpropane | C4H10 | 75-28-5 | 3001, 1003 |
| o-Xylene | C8H10 | 95-47-6 | 4009, 2011 |
| Propionitrile | C3H5N | 107-12-0 | 1001, 1041 |
| Furan | C4H4O | 110-00-9 | n.a. |
| 1-Chloropropane | C3H7Cl | 540-54-5 | 1001, 1002, 1044 |
| Di-n-propyl ether | C6H14O | 111-43-3 | 2001, 3002, 1025 |
| 1-Heptanol | C7H16O | 111-70-6 | 1001, 6002, 1014 |
| 1-Octanol | C8H18O | 111-87-5 | 1001, 7002, 1014 |
| Isopropylbenzene | C9H12 | 98-82-8 | 2001, 5009, 1013 |
| 1-Decene | C10H20 | 872-05-9 | 1001, 7002, 1005 |
| 3-Methyl-1-butene | C5H10 | 563-45-1 | 2001, 1003, 1005 |
| trans-1,3-Pentadiene | C5H8 | 2004-70-8 | 1001, 1005, 1006 |
| 2-Methyl-2-pentene | C6H12 | 625-27-4 | 3001, 1002, 1008 |
| 1,3-Butadiene | C4H6 | 106-99-0 | 2005 |
| 2,3-Dimethylbutane | C6H14 | 79-29-8 | 4001, 2003 |
| 1-Butene | C4H8 | 106-98-9 | 1001, 1002, 1005 |
| Propylbenzene | C9H12 | 103-65-1 | 1001, 1002, 5009, 1012 |
| 2,2-Dimethylbutane | C6H14 | 75-83-2 | 4001, 1002, 1004 |
| Ethyl butyrate | C6H12O2 | 105-54-4 | 2001, 2002, 1022 |
| Isobutyl acetate | C6H12O2 | 110-19-0 | 2001, 1002, 1003, 1021 |
| Acetic acid isopropyl ester | C5H10O2 | 108-21-4 | 2001, 1003, 1021 |
| Cycloheptane | C7H14 | 291-64-5 | 7002 |
| Cyclooctane | C8H16 | 292-64-8 | 8002 |
| 4-Isopropyltoluene | C10H14 | 99-87-6 | 2001, 4009, 1011, 1013 |
| Nonane | C9H20 | 111-84-2 | 2001, 7002 |
| Propanal | C3H6O | 123-38-6 | 1001, 1002, 1020 |
| Methyl propanoate | C4H8O2 | 554-12-1 | 2001, 1022 |
| Ethylcyclohexane | C8H16 | 1678-91-7 | 1001, 6002, 1003 |
| Hexanoic acid methyl ester | C7H14O2 | 106-70-7 | 2001, 3002, 1022 |
| Amyl acetate | C7H14O2 | 628-63-7 | 1001, 4002, 1021 |
| Diisobutyl ketone | C9H18O | 108-83-8 | 4001, 1002, 2003, 1019 |
| Formic acid propyl ester | C4H8O2 | 110-74-7 | 1023, 2002, 1001 |
| Methyl isopropyl ketone | C5H10O | 563-80-4 | 2001, 1003, 1018 |
| Isobutylene | C4H8 | 115-11-7 | 2001, 1007 |
| Perfluorohexane | C6F14 | 355-42-0 | 2074, 4075 |
| Biphenyl | C12H10 | 92-52-4 | 2010, 10009 |
| Eicosane | C20H42 | 112-95-8 | 2001, 18002 |
| 1,3,5-Trimethylbenzene | C9H12 | 108-67-8 | 3009, 3011 |
| Benzyl chloride | C7H7Cl | 100-44-7 | 5009, 1010, 1044 |
| Limonene | C10H16 | 138-86-3 | 2001, 3002, 1003, 1007, 1008 |
| Hexadecane | C16H34 | 544-76-3 | 2001, 14002 |
| Sulfolane | C4H8O2S | 126-33-0 | 2002, 1118 |
| 2,4,4-Trimethyl-1-pentene | C8H16 | 107-39-1 | 4001, 1002, 1004, 1007 |
| Diisobutyl ether | C8H18O | 628-55-7 | 4001, 1002, 2003, 1025 |
| 1-Hexyne | C6H10 | 693-02-7 | 1001, 3002, 1065 |
| 1-Heptyne | C7H12 | 628-71-7 | 1001, 4002, 1065 |
| 1-Heptene | C7H14 | 592-76-7 | 1001, 4002, 1005 |
| 1,5-Hexadiene | C6H10 | 592-42-7 | 2002, 2005 |
| 1-Pentyne | C5H8 | 627-19-0 | 1001, 2002, 1065 |
| 2-Hexanone | C6H12O | 591-78-6 | 1001, 3002, 1018 |
| o-Methylaniline | C7H9N | 95-53-4 | 4009, 1011, 1036 |
| Xylene | C8H10 | 1330-20-7 | 4009, 2011 |
| tert-Pentanol | C5H12O | 75-85-4 | 3001, 1002, 1004, 1014 |
| Dibromomethane | CH2Br2 | 74-95-3 | 1002, 2064 |
| Propyl bromide | C3H7Br | 106-94-5 | 2002, 1001, 1064 |
| Methyl butanoate | C5H10O2 | 623-42-7 | 2001, 1002, 1022 |
| n-Undecane | C11H24 | 1120-21-4 | 2001, 9002 |
| 2,3,4-Trimethyl pentane | C8H18 | 565-75-3 | 5001, 3003 |
| 1-Octyne | C8H14 | 629-05-0 | 1001, 5002, 1065 |
| Isopropyl bromide | C3H7Br | 75-26-3 | 2001, 1003, 1064 |
| Valeraldehyde | C5H10O | 110-62-3 | 1001, 3002, 1020 |
| Hexanal | C6H12O | 66-25-1 | 1001, 4002, 1020 |
| Octanal | C8H16O | 124-13-0 | 1001, 6002, 1020 |
| 2-Methylhexane | C7H16 | 591-76-4 | 3001, 3002, 1003 |
| Cycloheptatriene | C7H8 | 544-25-2 | 1002, 3006 |
| tert-Butylbenzene | C10H14 | 98-06-6 | 3001, 1004, 5009, 1010 |
| Tetrahydropyran | C5H10O | 142-68-7 | 4002, 1027 |
| Decalin | C10H18 | 91-17-8 | 8002, 2003 |
| o-Dichlorobenzene | C6H4Cl2 | 95-50-1 | 4009, 2053 |
| m-Methylaniline | C7H9N | 108-44-1 | 1011, 1036, 4009 |
| Methyl tert-butyl ether (MTBE) | C5H12O | 1634-04-4 | 3001, 1004, 1024 |
| Dipentyl ether | C10H22O | 693-65-2 | 1025, 2001, 7002 |
| Cyclopentene | C5H8 | 142-29-0 | 1006, 3002 |
| 1,4-Cyclohexadiene | C6H8 | 628-41-1 | 2006, 2002 |
| 4-Ethenylcyclohexene | C8H12 | 100-40-3 | 1005, 1006, 1003, 3002 |
| Methyl tert-amyl ether (TAME) | C6H14O | 994-05-8 | 1024, 3001, 1002, 1004 |
| Deuterium oxide | D2O | 7789-20-0 | 1016 |
| Hexyl acetate | C8H16O2 | 142-92-7 | 1001, 5002, 1021 |
| Methyl valerate | C6H12O2 | 624-24-8 | 2001, 2002, 1022 |
| Anthracene | C14H10 | 120-12-7 | 10009, 4010 |
| Phenanthrene | C14H10 | 85-01-8 | 10009, 4010 |
| 2-Octanol | C8H18O | 123-96-6 | 2001, 5002, 1003, 1014 |
| Butanenitrile | C4H7N | 109-74-0 | 1041, 1001, 1002 |
| cis-1,3-Pentadiene | C5H8 | 1574-41-0 | 1005, 1006, 1001 |
| Tetramethylstannane | C4H12Sn | 594-27-4 | n.a. |
| cis-2-Hexene | C6H12 | 7688-21-3 | 2001, 2002, 1006 |
| Carbon dioxide | CO2 | 124-38-9 | n.a. |
| 1,7-Octadiene | C8H14 | 3710-30-3 | 4002, 2005 |
| 2,2,2-Trifluoroethanol | C2H3F3O | 75-89-8 | 1002, 1014, 1074 |
| 3-Heptanone | C7H14O | 106-35-4 | 2001, 3002, 1019 |
| 2,2-Dimethylpentane | C7H16 | 590-35-2 | 4001, 2002, 1004 |
| trans-1,4-Dimethylcyclohexane | C8H16 | 2207-04-7 | 2001, 4002, 2003 |
| 1,3-Cyclohexadiene | C6H8 | 592-57-4 | 2002, 2006 |
| N,N-Dimethyl propanoic acid amide | C5H11NO | 758-96-3 | 1001, 1002, 1097 |
| Pentanenitrile | C5H9N | 110-59-8 | 1001, 2002, 1041 |
| 1-Octanamine | C8H19N | 111-86-4 | 1001, 6002, 1029 |
| Dimethyl sulfide | C2H6S | 75-18-3 | 1001, 1102 |
| p-Terphenyl | C18H14 | 92-94-4 | 14009, 4010 |
| Triacontane | C30H62 | 638-68-6 | 2001, 28002 |
| Isobutyronitrile | C4H7N | 78-82-0 | n.a. |
| 1-Aminopentane | C5H13N | 110-58-7 | 1001, 3002, 1029 |
| Dimethyl ethyl amine | C4H11N | 598-56-1 | 2001, 1002, 1034 |
| 1-Chloropentane | C5H11Cl | 543-59-9 | 1001, 3002, 1044 |
| 2,5-Dimethylhexane | C8H18 | 592-13-2 | 4001, 2002, 2003 |
| Iodobenzene | C6H5I | 591-50-4 | 5009, 1010, 1063 |
| Ethyl tert-butyl ether (ETBE) | C6H14O | 637-92-3 | 4001, 1004, 1025 |
| Chrysene | C18H12 | 218-01-9 | 12009, 6010 |
| Hexanenitrile | C6H11N | 628-73-9 | 1001, 3002, 1041 |
| 1-Phenyldodecane | C18H30 | 123-01-3 | 1001, 10002, 5009, 1012 |
| n-Butylcyclohexane | C10H20 | 1678-93-9 | 1001, 8002, 1003 |
| N-Methylcaprolactam | C7H13NO | 2556-73-2 | n.a. |
| trans-2-Pentene | C5H10 | 646-04-8 | 2001, 1002, 1006 |
| Heptylamine | C7H17N | 111-68-2 | 1001, 5002, 1029 |
| 1,3-Butadiene, 2,3-dimethyl- | C6H10 | 513-81-5 | 2001, 2007 |
| Benzyl bromide | C7H7Br | 100-39-0 | 5009, 1012, 1064 |
| 2,5-Dimethylpyrazine | C6H8N2 | 123-32-0 | n.a. |
| Tetraethylstannane | C8H20Sn | 597-64-8 | n.a. |
| 1-Octen-3-ol | C8H16O | 3391-86-4 | 1001, 4002, 1003, 1005, 1014 |
| 1-Octadecyl naphthalene | C28H44 | 26438-29-9 | 1001, 16002, 7009, 2010, 1012 |
| 1-Dodecyl | C22H42 | | 1001, 18002, 3003 |
| 1,2-Epoxy-p-menth-8-ene | C10H16O | 1195-92-2 | n.a. |

**Table S3.** Overview of the components that were considered as solvents in the present work. All information is adopted from the Dortmund Data Bank (DDB)[1]. In the last column, the group split according to modified UNIFAC (Dortmund)[2,3] is given, if applicable: the last three digits of each number define the subgroup, whereas with the first (two) digit(s) the count of the respective group per molecule is given.

| Component name | Chemical formula | CAS number | UNIFAC groups |
|---|---|---|---|
| Acetonitrile | C2H3N | 75-05-8 | 1040 |
| Acetone | C3H6O | 67-64-1 | 1001, 1018 |
| Ethylenediamine | C2H8N2 | 107-15-3 | 2029 |
| Ethyl bromide | C2H5Br | 74-96-4 | 1001, 1002, 1064 |
| 1,2-Ethanediol | C2H6O2 | 107-21-1 | 1062 |
| Ethanol | C2H6O | 64-17-5 | 1001, 1002, 1014 |
| Diethyl ether | C4H10O | 60-29-7 | 2001, 1002, 1025 |
| Aniline | C6H7N | 62-53-3 | 5009, 1036 |
| Methoxybenzene | C7H8O | 100-66-3 | 5009, 1010, 1024 |
| 2-Methylpyridine | C6H7N | 109-06-8 | 1001, 1038 |
| Ethyl acetate | C4H8O2 | 141-78-6 | 1001, 1002, 1021 |
| Benzyl alcohol | C7H8O | 100-51-6 | 5009, 1012, 1014 |
| Bromobenzene | C6H5Br | 108-86-1 | 5009, 1010, 1064 |
| Chlorobenzene | C6H5Cl | 108-90-7 | 5009, 1053 |
| Benzonitrile | C7H5N | 100-47-0 | n.a. |
| Nitrobenzene | C6H5NO2 | 98-95-3 | 5009, 1057 |
| Benzene | C6H6 | 71-43-2 | 6009 |
| 2-Butoxyethanol | C6H14O2 | 111-76-2 | 1001, 3002, 1100 |
| 1-Butanol | C4H10O | 71-36-3 | 1001, 3002, 1014 |
| 2-Butanone | C4H8O | 78-93-3 | 1001, 1002, 1018 |
| cis-1,2-Dichloroethylene | C2H2Cl2 | 156-59-2 | 1006, 2069 |
| 2-Chloroethanol | C2H5ClO | 107-07-3 | 1044, 1002, 1014 |
| Chloroform | CHCl3 | 67-66-3 | 1050 |
| 3-Methylphenol | C7H8O | 108-39-4 | 4009, 1011, 1017 |
| Cyclohexane | C6H12 | 110-82-7 | 6002 |
| Dibutyl ether | C8H18O | 142-96-1 | 2001, 5002, 1025 |
| Decane | C10H22 | 124-18-5 | 2001, 8002 |
| 1,1-Dichloroethane [R150a] | C2H4Cl2 | 75-34-3 | 1001, 1048 |
| 1,2-Dichloroethane | C2H4Cl2 | 107-06-2 | 2044 |
| trans-1,2-Dichloroethene | C2H2Cl2 | 156-60-5 | 1006, 2069 |
| Dichloromethane | CH2Cl2 | 75-09-2 | 1047 |
| N,N-Dimethylformamide | C3H7NO | 68-12-2 | 1072 |
| 1,4-Dioxane | C4H8O2 | 123-91-1 | 2002, 2027 |
| 2,4-Dimethylsulfolane | C6H12O2S | 1003-78-7 | 2001, 1002, 1003, 1119 |
| 2,6-Dimethylpyridine | C7H9N | 108-48-5 | 2001, 1039 |
| Dodecane | C12H26 | 112-40-3 | 2001  10002 |
| Butyl acetate | C6H12O2 | 123-86-4 | 1001, 3002, 1021 |
| Methyl acetate | C3H6O2 | 79-20-9 | 1001, 1021 |
| Acetic acid | C2H4O2 | 64-19-7 | 1001, 1042 |
| Furfural | C5H4O2 | 98-01-1 | 1061 |
| Hexane | C6H14 | 110-54-3 | 2001, 4002 |
| Heptane | C7H16 | 142-82-5 | 2001, 5002 |
| 2-Heptanone | C7H14O | 110-43-0 | 1001, 4002, 1018 |
| 2-Propanol | C3H8O | 67-63-0 | 2001, 1003, 1014 |
| Diisopropyl ether | C6H14O | 108-20-3 | 4001, 1003, 1026 |
| 2,2,4-Trimethylpentane | C8H18 | 540-84-1 | 5001, 1002, 1003, 1004 |
| 1-Hexene | C6H12 | 592-41-6 | 1001, 3002, 1005 |
| 1-Methylnaphthalene | C11H10 | 90-12-0 | 7009, 1011, 2010 |
| Methanol | CH4O | 67-56-1 | 1015 |
| 2-Methoxyethanol | C3H8O2 | 109-86-4 | 1001, 1100 |
| Nitromethane | CH3NO2 | 75-52-5 | 1054 |
| 1-Nitropropane | C3H7NO2 | 108-03-2 | 1001, 1002, 1055 |
| Octane | C8H18 | 111-65-9 | 2001, 6002 |
| 1-Octene | C8H16 | 111-66-0 | 1001, 5002, 1005 |
| Pentane | C5H12 | 109-66-0 | 2001, 3002 |
| 1-Pentanol | C5H12O | 71-41-0 | 1001, 4002, 1014 |
| 2-Pentanone | C5H10O | 107-87-9 | 1001, 2002, 1018 |
| Phenol | C6H6O | 108-95-2 | 5009, 1017 |
| 1-Propanol | C3H8O | 71-23-8 | 1001, 2002, 1014 |
| Pyridine | C5H5N | 110-86-1 | 1037 |
| Carbon disulfide | CS2 | 75-15-0 | 1058 |
| Dimethyl sulfoxide | C2H6OS | 67-68-5 | 1067 |
| tert-Butanol | C4H10O | 75-65-0 | 3001, 1004, 1014 |
| Tetradecane | C14H30 | 629-59-4 | 2001  12002 |
| trans-Decahydronaphthalene | C10H18 | 493-02-7 | 8002, 2003 |
| 1,2,3,4-Tetrahydronaphthalene | C10H12 | 119-64-2 | 2002, 4009, 2012 |
| Tetrachloromethane | CCl4 | 56-23-5 | 1052 |
| Tetrahydrofurfuryl alcohol | C5H10O2 | 97-99-4 | 3002, 1003, 1014, 1027 |
| Tetrahydrofuran | C4H8O | 109-99-9 | 3002, 1027 |
| Toluene | C7H8 | 108-88-3 | 5009, 1011 |
| Triethylamine | C6H15N | 121-44-8 | 3001, 2002, 1035 |
| 1,1,1-Trichloroethane [R140a] | C2H3Cl3 | 71-55-6 | 1001, 1051 |
| Water | H2O | 7732-18-5 | 1016 |
| p-Xylene | C8H10 | 106-42-3 | 4009, 2011 |
| Nitroethane | C2H5NO2 | 79-24-3 | 1001, 1055 |
| Cyclopentanol | C5H10O | 96-41-3 | 4002, 1003, 1014 |
| Fluorobenzene | C6H5F | 462-06-6 | 5009, 1071 |
| 1,1,2,2-Tetrachloroethane | C2H2Cl4 | 79-34-5 | 2048 |
| N-Methylformamide | C2H5NO | 123-39-7 | n.a. |
| N,N-Dimethylacetamide | C4H9NO | 127-19-5 | 1001, 1097 |
| Glycerol | C3H8O3 | 56-81-5 | 2002, 1003, 3014 |
| Propyl acetate | C5H10O2 | 109-60-4 | 1001, 2002, 1021 |
| Cyclopentanone | C5H8O | 120-92-3 | 3002, 1019 |
| Cyclohexanone | C6H10O | 108-94-1 | 4002, 1019 |
| Cyclohexanol | C6H12O | 108-93-0 | 5002, 1003, 1014 |
| Ricinoleic acid | C18H34O3 | 141-22-0 | 1001, 13002, 1042, 1006, 1014, 1003 |
| 3-Methyl-1-butanol | C5H12O | 123-51-3 | 2001, 2002, 1003, 1014 |
| 2-Ethoxyethanol | C4H10O2 | 110-80-5 | 1001, 1002, 1100 |
| Furfuryl alcohol | C5H6O2 | 98-00-0 | n.a. |
| 1,2-Propanediol | C3H8O2 | 57-55-6 | 1001, 1002, 1003, 2014 |
| N-Methyl-2-pyrrolidone | C5H9NO | 872-50-4 | 1085 |
| 3-Pentanone | C5H10O | 96-22-0 | 2001, 1002, 1019 |
| N-Methylacetamide | C3H7NO | 79-16-3 | 1001, 1095 |
| 1-Hexanol | C6H14O | 111-27-3 | 1001, 5002, 1014 |
| Hexafluorobenzene | C6F6 | 392-56-3 | 6071 |
| Perfluoro-n-heptane | C7F16 | 335-57-9 | 2074, 5075 |
| Perfluorotributylamine | C12F27N | 311-89-7 | n.a. |
| cis-Decahydronaphthalene | C10H18 | 493-01-6 | 8002, 2003 |
| Propionitrile | C3H5N | 107-12-0 | 1001, 1041 |
| 1-Heptanol | C7H16O | 111-70-6 | 1001, 6002, 1014 |
| 1-Octanol | C8H18O | 111-87-5 | 1001, 7002, 1014 |
| 1-Decene | C10H20 | 872-05-9 | 1001, 7002, 1005 |
| Ethyl butyrate | C6H12O2 | 105-54-4 | 2001, 2002, 1022 |
| Acetophenone | C8H8O | 98-86-2 | 5009, 1010, 1018 |
| Cycloheptanol | C7H14O | 502-41-0 | 6002, 1003, 1014 |
| Nonane | C9H20 | 111-84-2 | 2001, 7002 |
| Amyl acetate | C7H14O2 | 628-63-7 | 1001, 4002, 1021 |
| 1,4-Dicyanobutane | C6H8N2 | 111-69-3 | 2002, 2041 |
| Quinoline | C9H7N | 91-22-5 | 4009, 1039 |
| Phenylcyclohexane | C12H16 | 827-52-1 | 5002, 5009, 1013 |
| Triethylene glycol | C6H14O4 | 112-27-6 | 2002, 2100 |
| Chlorocyclohexane | C6H11Cl | 542-18-7 | 5002, 1045 |
| Diethylene glycol monomethyl ether | C5H12O3 | 111-77-3 | 2002, 1024, 1100 |
| 2-Isopropoxyethanol | C5H12O2 | 109-59-1 | 2001, 1003, 1100 |
| Diethylene glycol | C4H10O3 | 111-46-6 | 2002, 1014, 1100 |
| Perfluorohexane | C6F14 | 355-42-0 | 2074, 4075 |
| Acetic acid benzyl ester | C9H10O2 | 140-11-4 | 5009, 1012, 1021 |
| Diethylene glycol diethyl ether | C8H18O3 | 112-36-7 | 2001, 3002, 3025 |
| Octamethylcyclotetrasiloxane | C8H24O4Si4 | 556-67-2 | 8001, 4084 |
| Limonene | C10H16 | 138-86-3 | 2001, 3002, 1003, 1007, 1008 |
| Hexadecane | C16H34 | 544-76-3 | 2001 14002 |
| Phthalic acid dibutyl ester | C16H22O4 | 84-74-2 | 2001, 6002, 4009, 2010, 2077 |
| 1-Dodecanol | C12H26O | 112-53-8 | 1001 11002, 1014 |
| Sulfolane | C4H8O2S | 126-33-0 | 2002, 1118 |
| Monoethanolamine | C2H7NO | 141-43-5 | 1002, 1014, 1029 |
| 2,5-Hexanedione | C6H10O2 | 110-13-4 | 2002, 2018 |
| 1,1,1,3,3,3-Hexafluoro-2- | C3H2F6O | 920-66-1 | 1003, 1014, 2074 |
| Phthalic acid diethyl ester | C12H14O4 | 84-66-2 | 2001, 2002, 4009, 2010, 2077 |
| Tripentylamine | C15H33N | 621-77-2 | 3001, 11002, 1035 |
| Ethoxybenzene | C8H10O | 103-73-1 | 1001, 5009, 1010, 1025 |
| 1,4-Butanediol | C4H10O2 | 110-63-4 | 4002, 2014 |
| 3,3'-Oxybispropionitrile | C6H8N2O | 1656-48-0 | 1002, 1025, 2041 |
| gamma-Butyrolactone | C4H6O2 | 96-48-0 | 2002, 1022 |
| Bis(2-ethylhexyl) phthalate | C24H38O4 | 117-81-7 | 4001, 10002, 2003, 4009, 2010, 2077 |
| 1,1,2,2-Tetrabromoethane | C2H2Br4 | 79-27-6 | 2003, 4064 |
| Phthalic acid dinonyl ester | C26H42O4 | 84-76-4 | 2001, 16002, 4009, 2010, 2077 |
| Phthalic acid benzyl butyl ester | C19H20O4 | 85-68-7 | 1001, 3002, 9009, 2010, 1012, 2077 |
| Formamide | CH3NO | 75-12-7 | n.a. |
| Ethyl benzoate | C9H10O2 | 93-89-0 | 1001, 1002, 5009, 1010, 1077 |
| 1,5-Pentanediol | C5H12O2 | 111-29-5 | 5002, 2014 |
| Propylene carbonate | C4H6O3 | 108-32-7 | n.a. |
| 1,3-Propanediol | C3H8O2 | 504-63-2 | 3002, 2014 |
| 1,6-Hexanediol | C6H14O2 | 629-11-8 | 6002, 2014 |
| Dichloroacetic acid | C2H2Cl2O2 | 79-43-6 | 1042, 1048 |
| Indene | C9H8 | 95-13-6 | 1006, 4009, 1010, 1012 |
| 2,2'-Diethyl-dihydroxy sulfide | C4H10O2S | 111-48-8 | 3002, 2014, 1103 |
| Tetramethylene sulfoxide | C4H8OS | 1600-44-8 | n.a. |
| 2-Mercapto ethanol | C2H6OS | 60-24-2 | 1002, 1014, 1060 |
| Divinylsulfone | C4H6O2S | 77-77-0 | n.a. |
| 3-Methyl sulfolane | C5H10O2S | 872-93-5 | 1001, 1002, 1003, 1118 |
| 1,2-Dicyanoethane | C4H4N2 | 110-61-2 | 2041 |
| Decalin | C10H18 | 91-17-8 | 8002, 2003 |
| 2,4-Pentanedione | C5H8O2 | 123-54-6 | 1002, 2018 |
| Glutaronitrile | C5H6N2 | 544-13-8 | 2041, 1002 |
| Acetanilide | C8H9NO | 103-84-4 | n.a. |
| Methyl diphenyl phosphate | C13H13O4P | 115-89-9 | n.a. |
| Diethyl oxalate | C6H10O4 | 95-92-1 | 2001, 2002, 2077 |
| Deuterium oxide | D2O | 7789-20-0 | 1016 |
| Hexyl acetate | C8H16O2 | 142-92-7 | 1001, 5002, 1021 |
| Tributylamine | C12H27N | 102-82-9 | 3001, 8002, 1035 |
| Butanenitrile | C4H7N | 109-74-0 | 1041, 1001, 1002 |
| Dimethylcyanamide | C3H6N2 | 1467-79-4 | n.a. |
| Diiodomethane | CH2I2 | 75-11-6 | 1002, 2063 |
| Ethylene cyanohydrin | C3H5NO | 109-78-4 | 1002, 1014, 1041 |
| Squalane | C30H62 | 111-01-3 | 8001, 16002, 6003 |
| Benzylcyanide | C8H7N | 140-29-4 | 5009, 1010, 1041 |
| Phenylacetone | C9H10O | 103-79-7 | 5009, 1012, 1018 |
| 4-Phenyl-2-butanone | C10H12O | 2550-26-7 | 1002, 5009, 1012, 1018 |
| 2,2,2-Trifluoroethanol | C2H3F3O | 75-89-8 | 1002, 1014, 1074 |
| Trioctylamine | C24H51N | 1116-76-3 | 3001, 20002, 1035 |
| Bicyclohexyl | C12H22 | 92-51-3 | 10002, 2003 |
| N-Methyl propanamide | C4H9NO | 1187-58-2 | 1001, 1002, 1095 |
| N-Ethylacetamide | C4H9NO | 625-50-3 | 2001, 1096 |
| N,N-Dimethyl propanoic acid | C5H11NO | 758-96-3 | 1001, 1002, 1097 |
| Bromocyclohexane | C6H11Br | 108-85-0 | 5002, 1003, 1064 |
| Pentanenitrile | C5H9N | 110-59-8 | 1001, 2002, 1041 |
| Tributyl phosphate | C12H27O4P | 126-73-8 | n.a. |
| 2-Pyrrolidone | C4H7NO | 616-45-5 | n.a. |
| 1-Chloronaphthalene | C10H7Cl | 90-13-1 | 7009, 2010, 1053 |
| N-Formylmorpholine | C5H9NO2 | 4394-85-8 | n.a. |
| Bis-(2-ethylhexyl)-sebacate | C26H50O4 | 122-62-3 | 4001, 16002, 2003, 2022 |
| Trihexylamine | C18H39N | 102-86-3 | 3001, 14002, 1035 |
| alpha-Aminotoluene | C7H9N | 100-46-9 | 5009, 1010, 1029 |
| Hexamethylphosphoric acid triamide | C6H18N3OP | 680-31-9 | n.a. |
| Dimethyl ethyl amine | C4H11N | 598-56-1 | 2001, 1002, 1034 |
| Tetraethylene glycol | C8H18O5 | 112-60-7 | 3002, 1025, 2100 |
| Triethyl phosphate | C6H15O4P | 78-40-0 | n.a. |
| Trimethyl phosphate | C3H9O4P | 512-56-1 | n.a. |
| Octanenitrile | C8H15N | 124-12-9 | 1001, 5002, 1041 |
| Iodobenzene | C6H5I | 591-50-4 | 5009, 1010, 1063 |
| Ethyl tert-butyl ether (ETBE) | C6H14O | 637-92-3 | 4001, 1004, 1025 |
| Dibenzyl ether | C14H14O | 103-50-4 | 10009, 1010, 1012, 1025 |
| Bis(2-ethylhexyl) phosphate | C16H35O4P | 298-07-7 | n.a. |
| N-Acetyloxazolidine | C5H9NO2 | 3672-60-4 | 1001, 1027, 1099 |
| 2-Phenylethanol | C8H10O | 60-12-8 | 1002, 5009, 1012, 1014 |
| 1,5-Dimethyl-2-pyrrolidone | C6H11NO | 5075-92-3 | n.a. |
| 4-Chloromethyl-2-one-1,3-dioxolane | C4H5ClO3 | 2463-45-8 | n.a. |
| Hexanenitrile | C6H11N | 628-73-9 | 1001, 3002, 1041 |
| Heptanenitrile | C7H13N | 629-08-3 | 1001, 4002, 1041 |
| Nonanenitrile | C9H17N | 2243-27-8 | 1001, 6002, 1041 |
| 1,5-Dicyanopentane | C7H10N2 | 646-20-8 | 3002, 2041 |
| 1,6-Dicyanohexane | C8H12N2 | 629-40-3 | 4002, 2041 |
| Malonic acid dinitrile | C3H2N2 | 109-77-3 | n.a. |
| N-Acetylpiperidine | C7H13NO | 618-42-8 | 1001, 3002, 1099 |
| Carbonic acid diethyl ester | C5H10O3 | 105-58-8 | n.a. |
| Ethylene carbonate | C3H4O3 | 96-49-1 | n.a. |
| Ethylene sulfite | C2H4O3S | 3741-38-6 | n.a. |
| Ethyl phenyl ketone | C9H10O | 93-55-0 | 1001, 5009, 1010, 1019 |
| 4-Bromoanisole | C7H7BrO | 104-92-7 | 4009, 2010, 1024, 1064 |
| Di(2-ethylhexyl) adipate | C22H42O4 | 103-23-1 | 4001, 2003, 12002, 2022 |
| Pentadecanoic acid, nitrile | C15H29N | 18300-91-9 | 1001, 12002, 1041 |
| Cyclohexyl acetone | C9H16O | 103-78-6 | 6002, 1003, 1018 |
| Methylglutaronitrile | C6H8N2 | 4553-62-2 | n.a. |
| Methyleneglutaronitrile | C6H6N2 | 1572-52-7 | n.a. |
| beta-Chloropropionitrile | C3H4ClN | 542-76-7 | 1041, 1044 |
| N-Methylmethansulfonamide | C2H7NO2S | 1184-85-6 | n.a. |
| 1-Bromonaphthalene | C10H7Br | 90-11-9 | 7009, 3010, 1064 |
| N,N-Diethylacetamide | C6H13NO | 685-91-6 | 3001, 1099 |
| Iminodipropionitrile | C6H9N3 | 111-94-4 | 1002, 1032, 2041 |
| Mono-n-butyl phosphate | C4H11O4P | 1623-15-0 | n.a. |
| Tris-butoxyethyl phosphate | C18H39O7P | 78-51-3 | n.a. |
| Di-n-butyl phosphate | C8H19O4P | 107-66-4 | n.a. |
| N,N-Dibutyl-2-ethylhexylamide | C16H33NO | 5831-86-7 | 4001, 8002, 1003, 1099 |
| N,N-Dimethylisobutyramide | C6H13NO | 21678-37-5 | 2001, 1003, 1097 |
| N-Isopropylformamide | C4H9NO | 16741-46-1 | n.a. |
| N-Isopropylacetamide | C5H11NO | 1118-69-0 | n.a. |
| N-Methylisobutyramide | C5H11NO | 2675-88-9 | 2001, 1003, 1095 |
| N-Ethylpropionamide | C5H11NO | 5129-72-6 | 2001, 1002, 1096 |
| N-Methyl-2-piperidone | C6H11NO | 931-20-4 | n.a. |
| N-Methylcaprolactam | C7H13NO | 2556-73-2 | n.a. |
| Propyl phenyl ketone | C10H12O | 495-40-9 | 1001, 1002, 5009, 1010, 1019 |
| 1,3-Dimethylimidazolidin-2-one | C5H10N2O | 80-73-9 | n.a. |
| Tetraethylene glycol dimethyl ether | C10H22O5 | 143-24-8 | 5002, 2024, 3025 |
| Ethylene glycol monopropyl ether | C5H12O2 | 2807-30-9 | 1001, 2002, 1100 |
| 1,3-Dimethoxybenzene | C8H10O2 | 151-10-0 | 4009, 2010, 2024 |
| Fumaronitrile | C4H2N2 | 764-42-1 | n.a. |
| Maleonitrile | C4H2N2 | 928-53-0 | n.a. |
| Linoleic acid | C18H32O2 | 60-33-3 | 1001, 12002, 2006, 1042 |
| N,N-Dibutyl-2,2-dimethylbutanamide | C14H29NO | 126926-50-9 | 5001, 5002, 1004, 1099 |
| Perfluoro-n-octane | C8F18 | 307-34-6 | 2074, 6075 |
| 1-(1-Naphthalenyl)ethanone | C12H10O | 941-98-0 | 7009, 3010, 1018 |
| N,N-Diethyl dodecanamide | C16H33NO | 3352-87-2 | 3001, 10002, 1099 |
| N-Ethyl-2-pyrrolidone | C6H11NO | 2687-91-4 | n.a. |
| 1,2-Epoxy-p-menth-8-ene | C10H16O | 1195-92-2 | n.a. |
| Choline chloride | C5H14ClNO | 67-48-1 | n.a. |
| Dimethylsulfolane | C6H12O2S | n.a. | 2001, 2003, 1118 |
| Sulfolanylamine | C4H9NO2S | n.a. | 1002, 1030, 1118 |
| Propyl sulfolanyl ether | C7H14O3S | n.a. | 1001, 2002, 1003, 1025, 1118 |
| Decyl sulfolanyl ether | C14H28O3S | n.a. | 1001, 9002, 1003, 1025, 1118 |
| Methylsulfolane | C5H10O2S | n.a. | 1001, 1002, 1003, 1118 |